\documentclass[12pt,a4paper]{article} 
\pdfoutput=1

\usepackage[top=25mm,bottom=25mm,left=25mm,right=25mm]{geometry}
\usepackage[hidelinks]{hyperref}

\usepackage{bbm}
\usepackage{array}
\usepackage{booktabs,dcolumn}
\usepackage[table]{xcolor}
\usepackage{mathptmx}       % selects Times Roman as basic font
\usepackage{helvet}         % selects Helvetica as sans-serif font
\usepackage{courier}        % selects Courier as typewriter font
\usepackage{type1cm}        % activate if the above 3 fonts are
\usepackage{makeidx}         % allows index generation
\usepackage{graphicx}        % standard LaTeX graphics tool
\usepackage{multicol}        % used for the two-column index
\usepackage[bottom]{footmisc}% places footnotes at page bottom

\usepackage[flushleft]{threeparttable}
\usepackage{adjustbox}
\usepackage{natbib}
\usepackage{placeins}
\usepackage{url}                         % not available on your system

\PassOptionsToPackage{obeyspaces}{url}% ~ \usepackage[...,obeyspaces]{url}

\makeindex             % used for the subject index
\usepackage{amsfonts}
\usepackage{lipsum}
\usepackage{enumitem}

\setlist[itemize]{leftmargin=*}

\newcommand{\virg}[1]{``#1''}

\usepackage{caption}
\usepackage{setspace}
\usepackage{lscape}
\usepackage{arydshln}
\usepackage{float}
\usepackage{tabularx}
\usepackage{multirow}
\usepackage{amsmath,bm}

\usepackage{enumitem}
\usepackage{rotating}

\setlist[itemize]{leftmargin=*}

\newcolumntype{.}{D{.}{.}{-1}}
\makeatletter
\newcolumntype{B}[3]{>{\boldmath\DC@{#1}{#2}{#3}}c<{\DC@end}}
\makeatother

\newcommand\mc[1]{\multicolumn{1}{c}{#1}} % shorthand macro
\newcolumntype{d}{D{.}{.}{5.5}}           % alignment on decimal marker

\newcolumntype{p}{D{.}{}{5.5}}  % alignment on "." marker, che però non appare in tabella

\usepackage{amsfonts}
\usepackage{lipsum}
\usepackage{caption}
\usepackage{lscape}
\usepackage{booktabs}
\usepackage{tabularx}
\usepackage{multirow}
\usepackage[flushleft]{threeparttable}
\usepackage{adjustbox}
\usepackage{dcolumn}
\usepackage{siunitx}
\usepackage{arydshln}
\usepackage{subcaption}

\usepackage[nolists]{endfloat}

\usepackage{xpatch}
\usepackage{blindtext}
\makeatletter
\xpatchcmd{\titlepage}{\@restonecolfalse\newpage}{\@restonecolfalse}{}{}
\xpatchcmd{\endtitlepage}{\if@restonecol\twocolumn \else \newpage \fi}{\if@restonecol\twocolumn \else  \fi}{\typeout{success}}{\typeout{fail}}
\makeatother

\usepackage{etoolbox}
\patchcmd{\pprintMaketitle}
  {\fi\hrule}% the second rule
  {\fi\ifvoid\extrainfobox\else\unvbox\extrainfobox\par\vskip10pt\fi\hrule}
  {}{}

\newsavebox\extrainfobox
{}

\newcommand{\info}[2]{\mathcal{#1}_{#2}}

\DeclareMathOperator{\avar}{Avar}

\newcommand{\RGARCH}{$\mathsf{RGARCH}$}
\newcommand{\GARCH}{$\mathsf{GARCH}$}
\newcommand{\GJR}{$\mathsf{GJR}$}
\newcommand{\MEM}{$\mathsf{MEM}$}
\newcommand{\HAR}{$\mathsf{HAR}$}
\newcommand{\AHAR}{$\mathsf{AHAR}$}
\newcommand{\AMEM}{$\mathsf{AMEM}$}
\newcommand{\MEMs}{$\mathsf{Spline-MEM}$}
\newcommand{\MEMc}{$\mathsf{Component-MEM}$}
\newcommand{\MEMMIDAS}{$\mathsf{MEM-MIDAS}$}
\newcommand{\GM}{$\mathsf{GM}$}
\newcommand{\DAGM}{$\mathsf{DAGM}$}
\newcommand{\DMEMX}{$\mathsf{DMEM-X}$}

\newcommand{\DMEM}{$\mathsf{DMEM}$}
\newcommand{\src}{\textit{short}--run}
\newcommand{\lrc}{\textit{long}--run}

\usepackage{titling}
\providecommand{\keywords}[1]{\textbf{Keywords:} #1}

\title{Doubly Multiplicative Error Models with\\
Long-- and Short--run Components}% (and Mixed Frequency Data)}

\author{A. Amendola$^*$ \and V. Candila$^\dag$ \and F. Cipollini$^\ddag$ \and G.M. Gallo$^{\S}$}
\date{%
    $^{*}$Dept. of Economics and Statistics, University of Salerno, Italy,  \href{mailto:alamendola@unisa.it}{alamendola@unisa.it}\\%
    $^{\dag}$MEMOTEF Department, Sapienza University of Rome, Italy,  \href{mailto:vincenzo.candila@uniroma1.it}{vincenzo.candila@uniroma1.it\\
  $^\ddag$Dipartimento di Statistica \virg{G. Parenti}, University of Florence, Italy, \href{mailto:cipollini@disia.unifi.it}{fabrizio.cipollini@unifi.it}\\
  $^{\S}$Italian Court of Audits (Corte dei conti), and NYU in Florence, Italy, \href{mailto:giampiero.gallo@nyu.edu}{giampiero.gallo@nyu.edu}}\\[2ex]%
    \today
}

\begin{document}

\begin{titlingpage}

\maketitle

%\author[Amendola]{A. Amendola}
%\author[Candila]{V. Candila \corref{cor1}}
%\author[Cipollini]{F. Cipollini}
%\author[Gallo]{G.M. Gallo}
%
%
%\address[Amendola]{Dept. of Economics and Statistics, University of Salerno, Italy,  \href{mailto:alamendola@unisa.it}{alamendola@unisa.it}}
%
%\address[Candila]{MEMOTEF Department, Sapienza University of Rome, Rome, Italy,  \href{mailto:vincenzo.candila@uniroma1.it}{vincenzo.candila@uniroma1.it}}
%
%\address[Cipollini]{Dipartimento di Statistica \virg{G. Parenti}, University of Florence, Italy, \href{mailto:cipollini@disia.unifi.it}{fabrizio.cipollini@unifi.it}}
%
%\address[Gallo]{Italian Court of Audits (Corte dei conti), and NYU in Florence, Italy, \href{mailto:giampiero.gallo@nyu.edu}{giampiero.gallo@nyu.edu}}

\begin{abstract}
We suggest the Doubly Multiplicative Error class of models (\DMEM) for modeling and forecasting realized volatility, which combines two components accommodating low--, respectively, high--frequency features in the data. We derive the theoretical properties of the Maximum Likelihood and Generalized Method of Moments estimators. Two such models are then proposed, the \MEMc, which uses daily data for both components, and the \MEMMIDAS, which exploits the logic of MIxed--DAta Sampling (MIDAS).  The empirical application involves the S\&P 500, NASDAQ, FTSE 100 and Hang Seng indices: irrespective of the market, both \DMEM's outperform the \HAR\, and other relevant \GARCH--type models.
\end{abstract}


%\begin{keyword}
%Multiplicative error model; MIDAS; Volatility forecasting.
%\end{keyword}

%\begin{extrainfo}
%\emph{JEL classifications: C22, C51, C58.}
%\end{extrainfo}

\end{titlingpage}

\section{Introduction}

More than forty years have passed since Engle's pioneering work \citep{Engle:1982} on modeling the conditional variance as an autoregressive process of observable variables. GARCH-type models \citep{Bollerslev:1986} are still playing a significant role in the financial econometrics literature. This is mainly due to the fact that this class of models allows to reproduce several stylized facts, such as the persistence in the conditional second moments (volatility clustering) and, in its extensions, the possibility of taking into account the slow moving or state dependent average volatility level. This empirical regularity can be suitably accommodated assuming that the dynamic evolution of volatility is driven by two components, a high- and a low-frequency one, which combine additively or multiplicatively \citep[][offer a comprehensive survey of the contributions in this field]{Amado:Silva:Terasvirta:2019}. As a matter of fact, several suggestions exist in the GARCH literature to model the low frequency component. For instance, \cite{Hamilton:Susmel:1994} and \cite{Dueker:1997} consider a Markov Switching framework,  \cite{Amado:Terasvirta:2008} a Smooth Transition context, \cite{Mazur:Pipien:2012} and \cite{Engle:Rangel:2008} introduce deterministic functions in order to make the unconditional variance time-varying with high persistence. This latter contribution points to a relationship between a time--varying average level of volatility and macroeconomic events related to the business cycle: since the macro--variables are observed at a lower frequency than that of the asset returns, the MIxed--DAta Sampling (MIDAS) approach suggested by \cite{Ghysels:Sinko:Valkanov:2007} was extended to allow the real economy to influence financial volatility \citep[GARCH--MIDAS model][]{Engle:Ghysels:Sohn:2013, Conrad:Loch:2015}. Some extensions are available, such as the Double Asymmetric GARCH--MIDAS (DAGM) introduced by \cite{Amendola:Candila:Gallo:2019}, where a variable available at a low frequency drives the slow moving level of volatility and is allowed to have differentiated effects according to its sign, determining a local time--varying trend around which a GJR--GARCH  \citep[][\GJR]{Glosten:Jaganathan:Runkle:1993} describes the \src\ dynamics.\footnote{A similar approach was independently developed by \cite{Pan:Liu:2018}.}

Volatility modeling has encountered a tremendous boost by the availability of ultra-high frequency data, and the ensuing stream of literature related to estimating volatility using tick--by--tick data, conveniently aggregated: following the pathbreaking paper by \cite{Andersen:Bollerslev:1998}, realized volatility measures have become an ideal target for evaluating volatility forecasting performances. Such forecasts may be generated by GARCH models (for the conditional variances of asset returns) or by models of realized variances themselves (conditional expectations of variances or volatility, or, yet, log--variances), the latter being able to exploit intra-daily information about market movements. For the latter class of models a wide choice exists: the variants of the Multiplicative Error Model \citep[\MEM,][]{Engle:2002, Engle:Gallo:2006}, the Heterogeneous Autoregressive Model (\HAR) by \cite{Corsi:2009}, the Realized GARCH \citep[\RGARCH,][]{Hansen:Huang:Shek:2012}, among others, have proven to be effective in translating the refinement of volatility measurement achieved in the realized variance estimators \citep[for a survey on this estimators in reference to forecasting,  cf.][]{Andersen:Bollerslev:Christoffersen:Diebold:2006} into good out--of-sample model performances relative to the GARCH results (notoriously based just on squared close--to--close returns). 

This paper discusses the presence of a \lrc\ and a \src\ components of volatility, combining multiplicatively with one another within a unified general framework within the \MEM\ class, which we label \DMEM\ (Doubly Multiplicative Error Model): in it, the \src\ component is seen as fluctuating around one and be a function of past volatility or some predetermined variables, all observed at the same frequency. As per the \lrc\ component (which provides the time--varying average level of volatility), it can be assumed as: a constant (giving back the base \MEM); a smooth function of time (giving rise to a \MEMs\ in the case of a spline); a specification based on daily data which mirrors the structure of the \src\ component with a higher persistence (a novel model, which we label \MEMc); and the extension of the MIDAS approach to the \MEM world, providing a tool in which weekly or monthly data for the \lrc\ can be combined with daily data for the \src\ (another novel model, the \MEMMIDAS). From an empirical point of view we are motivated to compare performance of these models against a few representative models in the \GARCH\, class, in particular those based on a MIDAS approach on the one side and models for realized volatility keeping a base asymmetric MEM (\AMEM) as a reference, together with (an asymmetric versio of) the \HAR\, and the \RGARCH, all characterized by the absence of such a low--frequency component.

The theoretical discussion shows that both new models have desirable statistical properties for their estimators (both within a Maximum Likelihood and a Generalized Method of Moments framework). From an empirical point of view, we estimate all the competing models for the realized volatility series of four major indices (the S\&P 500, NASDAQ, FTSE 100 and Hang Seng). To summarize the results, to a question like \textit{Is a \lrc\ component advisable?}, the answer is yes: the models that do not use it are dominated by the ones that do within the classes of models for realized volatility on the one hand and models for conditional variances of returns on the other. To a question like \textit{Does modeling realized volatility perform better than a \GARCH, even when the latter contain a long--term component?}, our answer is still yes, pointing to the richness of intra--daily information over the consideration of just returns. Moreover, our results favor the \DMEM\, approach over the \HAR\, in spite of its capability of mimicking long memory features in the data.

Our contribution parallels a number of papers where the issue of a low-frequency component was taken into account. Within the \MEM\, context, has been estimated in several ways: through regime switching and smooth transition functions \cite{Gallo:Otranto:2015}, by deterministic splines \cite{Brownlees:Gallo:2010} or by a semi-non-parametric vector \MEM, where the low-frequency term affecting several assets is obtained non-parametrically \cite{Barigozzi+Brownless+Gallo+Veredas:2014}. A comparison with those models goes beyond the scope of this paper. 

The rest of the paper is organized as follows. In Section \ref{sect:MEM-MIDAS} we suggest the rationale and the notation for the \DMEM, introducing the two new models (\MEMc\, and \MEMMIDAS). Section \ref{sect:Inference} presents the theoretical results on the estimators' properties and statistical inference. Section \ref{sect:Empirical} introduces the market indices used in the empirical estimation, presents the results in terms of in--sample estimation and performs the main forecasting comparison across the competing models. Section \ref{sect:Concl} contains some concluding remarks.

\section{Multiplicative Error Models with Components}
\label{sect:MEM-MIDAS}

Let $\{x_{i,t}\}$ be a time series coming from a non-negative discrete time process for the $i$-th day ($i = 1, \ldots, N_t$) of the period $t$ (for example, a week, a month or a quarter; $t = 1 , \ldots, T$): this comprises most financial activity--related variables, such as realized volatility, high-low range, number of trades, volumes, durations, and so on. 

Let $\mathcal{F}_{i,t}$ be the information set available at day $i$ of period $t$. 
In its standard version \citep{Engle:2002}, the \MEM  assumes that
\begin{equation}
  \label{eqn:x-def}
  x_{i,t} = \mu_{i,t} \epsilon_{i,t} = \tau \xi_{i,t} \epsilon_{i,t},
\end{equation}
where: 
$\tau$ is a constant; $\xi_{i,t}$ is a quantity that, conditionally on $\mathcal{F}_{i-1,t}$ and by means of a parameter vector $\bm{\theta}$, evolves deterministically; 
$\epsilon_{i,t}$ is an error term such that 
\begin{equation}
  \label{eqn:eps-def}
  \epsilon_{i,t}|\info{F}{i-1,t} \overset{iid}{\sim} D^{+}(1, \sigma^2),
\end{equation}
meaning that it has a unit mean, unknown variance $\sigma^2$ and a probability density function defined over a non-negative support.% 
\footnote{For ease of notation, we use the set $\mathcal{F}_{i-1, t}$ even when the first day of a new period, say  $x_{1,t}$ depends on the information observed the last day of the period immediately preceding $t$, that is  $\mathcal{F}_{N_{t-1}, t-1}$}

Therefore, independently of the chosen distribution $D^+$ and the function used to build the evolution of $\mu_{i,t}$, we have that
\begin{equation}
 E(x_{i,t}|\mathcal{F}_{i-1,t})= \tau \xi_{i,t}.
 \label{condmean}
\end{equation}
Evaluating expression (\ref{condmean}) unconditionally, we can interpret $\tau$ to be the unconditional expectation of $x_{i,t}$ if we assume that $E(\xi_{i,t})=1$, so that $x_{i,t}$ moves around the constant term $\tau$. 
Correspondingly, the conditional variance can be expressed as
\begin{equation}
  Var(x_{i,t}|\mathcal{F}_{i-1,t})= \sigma^2 \tau^2 \xi_{i,t}^2.
\label{condvar}
\end{equation}

In this paper, we extend the specification for the conditional mean to have a multiplicative component structure, in which both factors of the conditional expectation are time--varying. 
We have
\begin{equation}
  \label{eqn:mu-mult}
  x_{i,t}= \mu_{i,t}\varepsilon_{i,t} = \tau_{i,t} \xi_{i,t} \varepsilon_{i,t}.
\end{equation}
$\tau_{i,t}$ can be seen as a \textit{slow}--moving component determining the average \textit{level} of the conditional mean at any given time, or, which is the same, a \lrc\, component. 
By the same token, since $\xi_{i,t}$ is a factor centered around one, it plays the role of dumping or amplifying $\tau_{i,t}$ depending on whether it is $<$ or $> 1$; for this reason, we label it as a \src\ or \textit{fast}--moving component.
Equation (\ref{eqn:mu-mult}) with innovation  (\ref{eqn:eps-def}) define a Doubly Multiplicative Error Model, or \DMEM.\footnote{The consideration of two multiplicative components in the univariate GARCH case is discussed by \cite{Conrad:Kleen:2020}.}

Let us start by expressing the \src\, component in general terms as the GARCH--type expression typical of a MEM, augmented by the contribution of a predetermined de--meaned (vector) variable $\bm{z}$ \citep[\DMEMX\, to parallel the $\mathsf{GARCH-X}$, cf. ][]{Han:Kristensen:2014}:
\begin{equation}
  \label{eqn:xi-1}
  \xi_{i,t} = \left(1 - \alpha_{1} - \gamma_{1} / 2 - \beta_{1} \right) + \alpha_{1} x^{(\xi)}_{i-1,t} + \gamma_{1} x^{(\xi-)}_{i-1,t} + \beta_{1} \xi_{i-1,t} + \bm{\delta}_1^\prime \bm{z}_{i-1, t}
\end{equation}
where
\begin{equation}
  \label{eqn:x^xi}
  x^{(\xi)}_{i,t}  \equiv \frac{x_{i,t}}{\tau_{i,t}} \qquad
  x^{(\xi-)}_{i,t} \equiv x^{(\xi)}_{i,t} \mathbbm{1}_{\left(r_{i,t}  < 0 \right)}.
\end{equation}
$x^{(\xi-)}_{i,t}$ is a variable derived from $x^{(\xi)}_{i,t}$ which takes a non-zero value only if it corresponds to a negative return (for asymmetric effects). 

Starting from $E(\xi_{i,t}) = 1$, we have
\begin{equation*}
E(\xi^{2}_{i,t}) = 
\frac{1 - \beta_{1}^{*2}}
{1 - \left[ (\sigma^{2} + 1) \left( (\beta_{1}^{*} - \beta_{1})^{2} + \gamma_{1}^{2} / 4 \right) + \beta_{1} (2 \beta_{1}^{*} - \beta_{1}) \right]}
\end{equation*}
where $\beta_{1}^{*} = \alpha_{1} + \gamma_{1} / 2 + \beta_{1}$ denotes the persistence. 
To simplify matters, here we removed the contribution of predetermined variables: explicit inclusion would require assumptions on the correlation between variables $x$ and $\bm{z}$. 
%The first leads to something similar to \citet[][p.4]{Conrad:Kleen:2020}. 

As far as the \lrc\, is concerned, we consider here different alternatives, apart from it being constant (the resulting model would be the standard \MEM).
\begin{itemize}
  \item[$\bullet$] [\MEMs] We can specify  $\tau_{i,t}$ by means of a spline function (for example a linear or a cubic spline)
  \begin{equation*}
    \tau_{i,t} = \exp(f_{s}(i,t))
  \end{equation*}
  as a smoothing spline or a regression spline with a relatively low number of knots so as to guarantee the \textit{slow}--moving feature.
  The resulting model is the so called \MEMs\, \citep[the P-Spline MEM of][corresponds to a specific choice of spline functions]{Brownlees:Gallo:2010}. \MEMs\, is trend-stationary (stationary around the trend component represented by $\tau_{i,t}$).

  \item[$\bullet$] [\MEMc] Another possibility is to structure $\tau_{i,t}$ in a way similar to $\xi_{i,t}$, namely
  \begin{equation*}
    \tau_{i,t} = \omega^{(\tau)} + \alpha_{1}^{(\tau)} x^{(\tau)}_{i-1,t} + \gamma_{1}^{(\tau)} x^{(\tau-)}_{i-1,t} + \beta_{1}^{(\tau)} \tau_{i-1,t}
  \end{equation*}
  where 
  \begin{equation}
    \label{eqn:x^tau}
    x^{(\tau)}_{i,t}  \equiv \frac{x_{i,t}}{\xi_{i,t}} \qquad
    x^{(\tau-)}_{i,t} \equiv x^{(\tau)}_{i,t} \mathbbm{1}_{\left(r_{i,t}  < 0 \right)}.
  \end{equation}
    
  The essential difference in comparison with $\xi_{i,t}$ is that $\tau_{i,t}$ is not constrained to move around a unit mean, although the persistence features of the components relative to one another characterize the fact that $\tau$ moves differently than $\xi$.   
  
  The model resulting from this specification of $\tau_{i,t}$, which we name \MEMc, is similar to the model introduced by~\citet{Brownlees:Cipollini:Gallo:2012} who  use, however, an additive (namely $\mu = \tau + \xi$) specification not examined here.
  Another specification which makes use of different multiplicative components is the Composite-\MEM\, proposed by \citet{Brownlees:Cipollini:Gallo:2011} to model intradaily volumes.
  
  The \MEMc\, is mean stationary $\Leftrightarrow$ 
  \begin{equation*}
    E \left( \tau_{i,t} \right) = 
    \mu 
    \left( 1 - 
    \frac{\sigma^{2} \left( \alpha_{1} + \gamma_{1} / 2 \right) \left( \alpha_{1}^{(\tau)} + \gamma_{1}^{(\tau)} / 2 \right) + 
      \left( \sigma^{2} + 1 \right) \gamma_{1}\gamma_{1}^{(\tau)}/4}
    {1 - \beta_{1}^{*} \beta_{1}^{(\tau)*}} 
    \right)
  \end{equation*}
  where $\mu = E(x_{i,t})$ and $\beta_{1}^{(\tau)*} = \alpha_{1}^{(\tau)} + \gamma_{1}^{(\tau)} / 2 + \beta_{1}^{(\tau)}$.
  If all parameters are non-negative, this implies that $E(\tau_{i,t}) \leq \mu$.
  Such characteristic comes from the fact that the drivers of $\xi$ and $\tau$ equations, namely $x_{i,t}^{(\xi)}$ and $x_{i,t}^{(\tau)}$, are positively correlated since they both depend on $\varepsilon_{i,t}$. 
  In case of mean-stationarity we have then
  \begin{equation*}
    \omega^{(\tau)} = 
    \mu 
    \left( 1 - pers(\tau) \right)
    \left( 1 - 
    \frac{\sigma^{2} \left( \alpha_{1} + \gamma_{1} / 2 \right) \left( \alpha_{1}^{(\tau)} + \gamma_{1}^{(\tau)} / 2 \right) + 
      \left( \sigma^{2} + 1 \right) \gamma_{1}\gamma_{1}^{(\tau)}/4}
    {1 - \beta_{1^{*}} \beta_{1}^{(\tau)*}} 
    \right)
  \end{equation*}
  Easier to understand in case $\gamma_{1} = \gamma_{1}^{(\tau)} = 0$: 
  \begin{equation*}
    E \left( \tau_{i,t} \right) = 
    \mu 
    \left( 1 - 
    \frac{\sigma^{2} \alpha_{1} \alpha_{1}^{(\tau)}}
    {1 - pers(\xi) pers(\tau)} 
    \right)
    \end{equation*}
    \begin{equation*}
    \omega^{(\tau)} = 
    \mu 
    \left( 1 - \beta_{1}^{*} \right)
    \left( 1 - 
    \frac{\sigma^{2} \alpha_{1} \alpha_{1}^{(\tau)}}
    {1 - \beta_{1}^{*} \beta_{1}^{(\tau)*}} 
    \right)
  \end{equation*}

  \item[$\bullet$] [\MEMMIDAS] Yet another option is to allow $\tau_{i,t}$ to have a MIDAS-like structure, adapting the use of mixed frequency data models \citep{Engle:Ghysels:Sohn:2013, Conrad:Kleen:2020} to the multiplicative error model context. 
  In its simplest form, for all days $i$ of the same period $t$, $\tau_{i, t}$ can be expressed over a window of $K$ periods as 
  \begin{equation*}
    \tau_{i, t} = \tau_{t} \equiv \exp \left\{ m + \zeta \sum_{j=1}^K \delta_{j}(\omega) X_{t-j}\right\}
  \end{equation*}
  where $X_{t}$ indicates a variable available only at $t$ times and 
  \begin{equation}
    \label{eq:beta}
    \delta_k(\omega) = \frac{(k/K)^{\omega_1-1} (1-k/K)^{\omega_2-1}}{\displaystyle \sum_{j=1}^K (j/K)^{\omega_1-1}(1-j/K)^{\omega_2-1}}.
  \end{equation}
  Assuming $\omega_1 = 1$ and $\omega_2 \geq 1$ in (\ref{eq:beta}) identifies cases in which more emphasis is given to most recent observations. 
  %\QQQFC{\citet{Conrad:Kleen:2020} la fanno pi\`u lunga\ldots}

A further refinement is inspired by the \DAGM\, \citep{Pan:Liu:2018, Amendola:Candila:Gallo:2019}.

Regarding the choice of the MIDAS driver $X$, one could favor a variable $X \perp \varepsilon$ as in \citet{Conrad:Kleen:2020}, as this simplifies the analysis, although it may be difficult to meet this condition in practice \citep[as acknowledged by][p.4]{Conrad:Kleen:2020}).

 % \item \QQQFC{Ci frega qualcosa di un MS-DMEM parallelo a MS-GARCH? \citet[][p.5]{Conrad:Kleen:2020}} nelle conclusioni
 
\end{itemize}

%  I have still unclear how to measure persistence in a \DMEM.  The reason is that $\mu_{t}$ depends on $\mu_{t-1}$ but also on $\tau_{t-1}$ and $\xi_{t-1}$.

% So far, the parameter space of the proposed \MEMMIDAS\, consists of $\Theta^{MEM-M}=\left\lbrace \alpha, \beta, m, \theta, \omega_2 \right\rbrace$, while $\mu$ can be replaced by its sample mean $\bar{x}$. However, the model (as the base MEM) is enough flexible to include some asymmetric terms both in the short- and long-run equations (as in \cite{amendola2019asymmetric}, for instance) and/or some additional lags of the variable $x_{i,t}$. Suppose that $x_{i,t}$ is a volatility measure, such the realized volatility. If the interest is on evaluating the impact of negative lagged daily returns $r_{i-1,t}$ on the today's volatility, the short-run equation of the \MEMMIDAS\, transforms to:
% \begin{equation}
% \mu_{i,t}  = (1-\alpha-\beta-\gamma/2) + \left(\alpha +  \gamma \cdot \mathbbm{1}_{\left(r_{i-1,t}  < 0 \right)}\right) \frac{x_{i-1,t}}{\tau_t} + \beta \mu_{i-1,t}.
% \end{equation}

%%%%%%%%%%%%%%%%%%%%
\section[Inference]{Inference}
\label{sect:Inference}
 
Inference on the model defined in Section~\ref{sect:MEM-MIDAS} can be obtained extending the framework suggested by \citet[][Section~9.2.2]{Brownlees:Cipollini:Gallo:2012}. 
Assuming that the conditional mean is correctly specified 
%\QQQFC{Probabilmente basta che il modello specificato includa la specificazione corretta; forse ci sarebbero forse da aggiungere altre technicalities (di quelle tipo Econometric Theory) ma per ora lascerei perdere} 
and indicating with $\bm{\theta}$ the vector of parameters entering it,  
two estimation strategies are illustrated in what follows: Maximum Likelihood (ML) and Generalized Method of Moments (GMM).

\subsection{Maximum Likelihood Inference}
\label{sect:MLInference}

The \DMEM\, Maximum Likelihood estimator $\widehat{\bm{\theta}}_{ML}$ is defined as the value of $\bm{\theta}$ maximizing the average \textit{log-likelihood} function 
\begin{equation*}
  \overline{l}_N
    =
  N^{-1} \sum_{t = 1}^T \sum_{i = 1}^{N_{t}} l_{i,t}
    =
  N^{-1} \sum_{t = 1}^T \sum_{i = 1}^{N_{t}} \left[ \ln f_\varepsilon( \varepsilon_{i,t} | \info{F}{i-1, t}) + \ln \varepsilon_{i,t} - \ln x_{i,t} \right]
\end{equation*}
where $\displaystyle N = \sum_{t = 1}^{T} N_{t}$ is the number of observations.
The portion relative to $\bm{\theta}$ of the average \textit{score} function can be expressed as
\begin{equation}
  \label{eqn:Score}
  \overline{\bm{s}}_N
    =
  N^{-1} \sum_{t = 1}^T \sum_{i = 1}^{N_{t}} \nabla_{\bm{\theta}} l_{i,t}
    =
  -N^{-1} \sum_{t = 1}^T \sum_{i = 1}^{N_{t}} (\varepsilon_{i,t} b_{i,t} + 1) \bm{a}_{i,t},
\end{equation}
where
\begin{align}
  \label{eqn:eps}
  \varepsilon_{i,t} & = \frac{x_{i,t}}{\tau_{i,t} \xi_{i,t}}
  \\
  \nonumber
  \bm{a}_{i,t} & = \frac{1}{\mu_{i,t}} \nabla_{\bm{\theta}} \mu_{i,t} = \frac{1}{\tau_{i,t}} \nabla_{\bm{\theta}} \tau_{i,t} + \frac{1}{\xi_{i,t}} \nabla_{\bm{\theta}} \xi_{i,t}
    \\
  \nonumber  
  b_{i,t} & = \nabla_{\varepsilon_{i,t}} \ln f_\varepsilon(\varepsilon_{i,t} | \info{F}{i-1,t}).
\end{align}
A choice of $f_\varepsilon( \varepsilon_{i,t} | \info{F}{i-1, t})$ giving 
\begin{equation}
  \label{eqn:Cond-1}
  E\left( \varepsilon_{i,t} b_{i,t} + 1 | \info{F}{i-1,t} \right) = 0
\end{equation} 
implies a zero expected score and, so, consistency of $\widehat{\bm{\theta}}_{ML}$.
This condition is obtained in case of correct specification of the error distribution but, as discussed in what follows, there are choices of $f_\varepsilon( \varepsilon_{i,t} | \info{F}{i-1, t})$ able to guarantee (\ref{eqn:Cond-1}) despite they are wrongly specified:
in this case, $\widehat{\bm{\theta}}_{ML}$ is said a QML estimator. 
In what follows we assume that (\ref{eqn:Cond-1}) is satisfied by the distribution chosen for $\varepsilon_{i,t}$.
 
The squared portions relative to $\bm{\theta}$ of the asymptotic \textit{OPG} ($\overline{\bm{I}}_\infty$) and \textit{Hessian} ($\overline{\bm{H}}_\infty$) matrices are given by $\lim_{N \rightarrow \infty}$ of, respectively, 
\begin{eqnarray}
  \label{eqn:OPGN}
  \overline{\bm{I}}_N
    &=&
      N^{-1} \sum_{t = 1}^T \sum_{i = 1}^{N_{t}} 
      E \left( \nabla_{\bm{\theta}} l_{i,t} \nabla_{\bm{\theta}^\prime} l_{i,t} \right) 
    =
  N^{-1} \sum_{t = 1}^T \sum_{i = 1}^{N_{t}} 
  E \left[ 
  \left( \varepsilon_{i,t} b_{i,t} + 1 \right)^2 | \info{F}{i-1, t}\right] 
  E \left( \bm{a}_{i,t}  \bm{a}_{i,t}^\prime \right)
  \\
  \nonumber
  \overline{\bm{H}}_N
    &=&
       N^{-1} \sum_{t = 1}^T \sum_{i = 1}^{N_{t}} E \left( \nabla_{\bm{\theta} \bm{\theta}^\prime} l_{i,t} \right)
  \\
  \nonumber
    &=&
  N^{-1} \sum_{t = 1}^T \sum_{i = 1}^{N_{t}} 
    \left[
    E \left[\varepsilon_{i,t} \left( b_{i,t} + \varepsilon_{i,t} \nabla_{\varepsilon_{i,t}} b_{i,t} \right) | \info{F}{i-1, t} \right]
    E \left( \bm{a}_{i,t}  \bm{a}_{i,t}^\prime \right)
    - E \left( \varepsilon_{i,t} b_{i,t} + 1 | \info{F}{i-1, t} \right)  
      E\left( \nabla_{\bm{\theta}} \bm{a}_{i,t}^\prime \right){}
    \right]
    \\
    &=&
  \label{eqn:HessianN}
  N^{-1} \sum_{t = 1}^T \sum_{i = 1}^{N_{t}} 
      E \left[\varepsilon_{i,t} \left( b_{i,t} + \varepsilon_{i,t} \nabla_{\varepsilon_{i,t}} b_{i,t} \right) | \info{F}{i-1, t} \right] 
    E \left( \bm{a}_{i,t}  \bm{a}_{i,t}^\prime \right)
\end{eqnarray}
where the last equality is implied by (\ref{eqn:Cond-1}).

Expressions (\ref{eqn:OPGN}) and (\ref{eqn:HessianN}) are sufficient to derive $\avar(\widehat{\bm{\theta}}_{ML})$ (the asymptotic variance matrix of $\widehat{\bm{\theta}}_{ML}$), but only when the possible free shape parameter in $f_\varepsilon( \varepsilon_{i,t} | \info{F}{i-1, t})$, say $\lambda$, is ``orthogonal'' to $\bm{\theta}$ in the sense that it satisfies
\begin{equation*}
 \lim_{N \rightarrow \infty} 
 \left[
  N^{-1} \sum_{t = 1}^T \sum_{i = 1}^{N_{t}} 
  E \left( \nabla_{\lambda} \nabla_{\bm{\theta}^{\prime}} l_{i,t} \right) 
  \right]
  =
 - \lim_{N \rightarrow \infty} 
 \left[
  N^{-1} \sum_{t = 1}^T \sum_{i = 1}^{N_{t}} 
  E \left( \varepsilon_{i,t} \nabla_{\lambda} b_{i,t} | \info{F}{i-1,t} \right) E \left( \bm{a}_{i,t} \right)   
  \right]
  = 
  \bm{0};
\end{equation*}
if this not happens, the variance matrix of $\widehat{\bm{\theta}}_{ML}$ depends also on the asymptotic variance of $\widehat{\lambda}$.%
\footnote{\label{fn:H12} Expressing the full parameter vector as $(\bm{\theta}; \lambda)$, the corresponding \textit{OPG} and \textit{Hessian} matrices are structured in $(i,j)$-blocks $(i,j = 1, 2)$ corresponding to the two parameters in that order.
Since $\avar(\widehat{\bm{\theta}}_{ML})$ is related to the $(1,1)$-block of some inverse matrix (being it the asymptotic \textit{OPG}, \textit{Hessian} or Sandwich matrix), in general it may depend on the asymptotic variance of $\widehat{\lambda}$, right as a consequence of the block matrix algebra.

For example, in case of correct model specification, 
\begin{equation*}
  \avar(\widehat{\bm{\theta}}_{ML}) = -(\overline{\bm{H}}_{11} - \overline{\bm{H}}_{12}\overline{\bm{H}}_{22}^{-1}\overline{\bm{H}}_{21})^{-1}
\end{equation*}
simplifies to $\avar(\widehat{\bm{\theta}}_{ML}) = \overline{\bm{H}}_{11}^{-1}$ only in case $\overline{\bm{H}}_{12} = \bm{0}$ (for sake of simplicity, we use symbols $\bm{I}$ and $\bm{H}$, reserved in this section to the parameter $\bm{\theta}$, also for the general $(\bm{\theta}; \lambda)$ case; we also omit the $\infty$ symbol).

If one refers instead the Sandwich matrix, we have in general 
\begin{equation*}
  \avar(\widehat{\bm{\theta}}_{ML}) = 
  \bm{A}^{-1} \left( 
  \overline{\bm{I}}_{11} - 
  \bm{B} \overline{\bm{I}}_{21}- \overline{\bm{I}}_{12} \bm{B}^{\prime} + \bm{B} \overline{\bm{I}}_{22} \bm{B}^{\prime} 
  \right) \bm{A}^{-1},
\end{equation*}
($\bm{A} = \overline{\bm{H}}_{11} - \overline{\bm{H}}_{12}\overline{\bm{H}}_{22}^{-1}\overline{\bm{H}}_{21}$ and  
$\bm{B} = \overline{\bm{H}}_{12} \overline{\bm{H}}_{22}^{-1}$) 
that simplifies to $\avar(\widehat{\bm{\theta}}_{ML}) = \overline{\bm{H}}_{11}^{-1} \overline{\bm{I}}_{11} \overline{\bm{H}}_{11}^{-1}$ again in case $\overline{\bm{H}}_{12} = \bm{0}$.
 
$\overline{\bm{H}}_{12} = \bm{0}$ is what is labeled ``orthogonality'' condition in the text. 
See \citet[][Section~6]{Newey:McFadden:1994} for a related discussion.}
Note that this ``orthogonality'' condition is trivially implied by
\begin{equation}
  \label{eqn:Ortho}
  E \left( \varepsilon_{i,t} \nabla_{\lambda} b_{i,t} | \info{F}{i-1,t} \right) = 0.
\end{equation}

In the following section we discuss two among the possible specifications of the error distribution.

%% The Gamma(phi, phi) case
\subsubsection{Gamma Error Distribution}

A sensible specification for the conditional distribution of $\varepsilon_{i,t}$ is the $Gamma(\phi, \phi)$, which guarantees the constraint $E(\varepsilon_{i,t}|\info{F}{i-1,t}) = 1$ and implies $V(\varepsilon_{i,t}|\info{F}{i-1,t}) = 1 / \phi$. 
This can be seen as a generalization introduced by \cite{Engle:Gallo:2006} to the choice of exponential distribution (where $\phi=1$) within the Autoregressive Conditional Durations (ACD) model by \cite{Engle:Russell:1998} and of the $\chi^2(1)$ distribution (where $\phi=2$) suggested by \cite{Engle:2002}.
In such a case,
\begin{equation}
  \label{eqn:b:Gamma}
  b_{i,t} =  \frac{\phi - 1}{\varepsilon_{i,t}} - \phi
    \qquad
  \Rightarrow
    \qquad
  \varepsilon_{i,t} b_{i,t} + 1 = \phi (1 - \varepsilon_{i,t}).
\end{equation}
It is important to remark that this choice guarantees condition (\ref{eqn:Cond-1}) is satisfied should the Gamma not be the true distribution of the error term (QML property), and irrespective of the value of $\phi$: this makes the results based on assuming the exponential or the $\chi^{2}(1)$ distributions much more general, upon an appropriate choice of the standard errors.

Plugging Equation~(\ref{eqn:b:Gamma}) into (\ref{eqn:Score}) provides the $\bm{\theta}$--portion of the average score
\begin{equation}
  \label{eqn:Score:Gamma}
  \overline{\bm{s}}_N
    =
  \phi N^{-1} \sum_{t = 1}^T \sum_{i = 1}^{N_{t}} (\varepsilon_{i,t} - 1) \bm{a}_{i,t},
\end{equation}
which, in turn, implies the first order condition
\begin{equation}
  \label{eqn:Score:Equation:Gamma}
  \sum_{t = 1}^T \sum_{i = 1}^{N_{t}} ( \varepsilon_{i,t} - 1 ) \bm{a}_{i,t} = \bm{0}.
\end{equation}

Equation~(\ref{eqn:b:Gamma}) guarantees also the important implication that the shape parameter $\phi$ is ``orthogonal'' to $\bm{\theta}$ in the sense of Equation~(\ref{eqn:Ortho}):
\begin{equation*}
  E \left( \varepsilon_{i,t} \nabla_{\phi} b_{i,t} | \info{F}{i-1,t} \right){}
  = 
  E \left( \varepsilon_{i,t} \left( \varepsilon_{i, t}^{-1} - 1 \right) | \info{F}{i,t} \right){}
  =
  0,
\end{equation*}
as a consequence of the unit mean assumption for the error term.
This, in turn, implies that the asymptotic variance of $\widehat{\bm{\theta}}_{ML}$ is uniquely determined by the \textit{OPG} and the \textit{Hessian} matrices 
\begin{equation*}
  \overline{\bm{I}}_\infty = \phi^2 \sigma^2 \bm{A}
  \qquad
  \overline{\bm{H}}_\infty = - \phi \bm{A},
\end{equation*}
where
\begin{equation*}
  \bm{A} = \lim_{N \rightarrow \infty} \left[ N^{-1} \sum_{t = 1}^T \sum_{i = 1}^{N_{t}} E \left( \bm{a}_{i,t}  \bm{a}_{i,t}^\prime \right) \right].
\end{equation*}
Correspondingly, the OPG, Hessian and Sandwich versions of the asymptotic variance matrix are, respectively,
\begin{align}
  \nonumber
  \avar_I(\widehat{\bm{\theta}}_{ML}) & =  \phi^{-2} \sigma^{-2} \bm{A}^{-1}
    \\
  \nonumber
  \avar_H(\widehat{\bm{\theta}}_{ML}) & =  \phi^{-1} \bm{A}^{-1}
    \\
  \label{eqn:Avar:Sandwich}
  \avar_S(\widehat{\bm{\theta}}_{ML}) & =  \sigma^{2} \bm{A}^{-1}.
\end{align}
Equivalence among the three expressions is ensured by taking $\phi = \sigma^{-2}$ (instead of fixing it, like for instance in the exponential and $\chi_2(1)$ cases); hence, a consistent estimator is
\begin{equation*}
  \widehat{\avar}(\widehat{\bm{\theta}}_{ML}) =
  \widehat{\sigma}^2 \widehat{\bm{A}}^{-1}
\end{equation*}
where $\widehat{\sigma}^2$ is a consistent estimator of $\sigma^2$, 
\begin{equation*}
  \widehat{\bm{A}} = 
  N^{-1} \sum_{t = 1}^T \sum_{i = 1}^{N_{t}} \widehat{\bm{a}}_{i,t} \widehat{\bm{a}}_{i,t}^\prime,
\end{equation*}
and $\widehat{\bm{a}}_{i,t}$ means $\bm{a}_{i,t}$ evaluated at $\widehat{\bm{\theta}}_{ML}$.

The ML estimator of $\phi$ solves
\begin{equation}
  \label{eqn:phiML-1}
  \ln \phi + 1 - \psi(\phi)  +  N^{-1} \sum_{t = 1}^T \sum_{i = 1}^{N_{t}} \left[\ln \widehat{\varepsilon}_{i,t} -  \widehat{\varepsilon}_{i,t} \right] = 0,
\end{equation}
where, $\psi(\cdot)$ denotes the \emph{digamma} function and $\widehat{\varepsilon}_{i,t}$ indicates the RHS of (\ref{eqn:eps}) where the denominator is evaluated at $\widehat{\bm{\theta}}_{ML}$.%
\footnote{Considering the unit expectation constraint on $\varepsilon_{i,t}$, we likely have $N^{-1} \sum_{t = 1}^T \sum_{i = 1}^{N_{t}} \widehat{\varepsilon}_{i,t} \approx 1$, so that (\ref{eqn:phiML-1}) could be simplified as
\begin{equation*}
  \ln \phi - \psi(\phi)  +  N^{-1} \sum_{t = 1}^T \sum_{i = 1}^{N_{t}} \ln \widehat{\varepsilon}_{i,t}  = 0.
\end{equation*}
} 
Of course, this estimator is efficient if the true distribution is Gamma, but it is unfeasible if zeros are present in the data, given that $\ln \varepsilon_{i,t} = \ln x_{i,t} - \ln \tau_{i,t} - \ln \xi_{i,t}$.
An alternative, which is not suffering from this drawback, is provided by using a GMM estimator of $\sigma^2$ (discussed below).

%% The Lognormal(-V/2, V) case
\subsubsection{Log-Normal Error Distribution}

Another possible specification for the conditional distribution of $\varepsilon_{i,t}$ is the $Lognormal(-V/2, V)$, which guarantees the constraint $E(\varepsilon_{i,t}|\info{F}{i-1,t}) = 1$ and implies $Var(\varepsilon_{i,t}|\info{F}{i-1,t}) = \exp(V) - 1$), assuming no zeros are present in the data.
In such case,
\begin{equation}
  \label{eqn:b:LogN}
  b_{i,t} =  -\frac{1}{\varepsilon_{i,t}} \left( 1.5 + \frac{\ln \varepsilon_{i,t}}{V} \right)
    \qquad
  \Rightarrow
    \qquad
  \varepsilon_{i,t} b_{i,t} + 1 = -V^{-1} \left( \frac{V}{2} + \ln \varepsilon_{i,t} \right).
\end{equation}
As noted before, if the Log-normal is the true distribution of $\varepsilon_{i,t}$ 
% (as we assume in what follows) 
then condition (\ref{eqn:Cond-1}) is satisfied; 
otherwise, this condition requires $E \left( \ln \varepsilon_{i,t} | \info{F}{i-1, t}\right) = -V/2$.
 
The resulting $\bm{\theta}$--portion of the average score is then given by
\begin{equation}
  \label{eqn:Score:LogN}
  \overline{\bm{s}}_N
    =
  V^{-1} N^{-1} \sum_{t = 1}^T \sum_{i = 1}^{N_{t}} \left( \ln \varepsilon_{i,t} + \frac{V}{2} \right) \bm{a}_{i,t},
\end{equation}
for the first order condition 
\begin{equation}
  \label{eqn:Score:Equation:LogN}
  \sum_{t = 1}^T \sum_{i = 1}^{N_{t}} \left( \ln \varepsilon_{i,t} + \frac{V}{2} \right) \bm{a}_{i,t} = \bm{0}.
\end{equation}
Notice that, differently from the Gamma case (cf.~Equation~(\ref{eqn:Score:Equation:Gamma})), Equation (\ref{eqn:Score:Equation:LogN}) depends on the shape parameter $V$. 
This implies that, during estimation, one should alternate between estimation of $\bm{\theta}$ and $V$.%

Another important difference with the Gamma case is that the shape parameter $V$ is not ``orthogonal'' to $\bm{\theta}$, given that the LHS of Equation~(\ref{eqn:Ortho}) is now
\begin{equation}
 \label{eqn:Ortho:V}
  E \left( \varepsilon_{i,t} \nabla_{V} b_{i,t} | \info{F}{i-1,t} \right){}
  = 
  E \left( \varepsilon_{i,t} \varepsilon_{i, t}^{-1} \ln \varepsilon_{i, t} V^{-2} | \info{F}{i-1,t} \right){}
  =
  -\frac{V}{2} V^{-2} 
  =
  -\frac{V^{-1}}{2};
\end{equation}
this implies that $\avar( \widehat{\bm{\theta}}_{ML})$ depends both on $V$ and on the asymptotic variance of an estimator $\widehat{V}$ (more on this below).

Focusing now on the shape parameter, the ML estimator of $V$ solves
\begin{equation}
  \label{eqn:VML}
  \frac{V^{2}}{4} + V - N^{-1} \sum_{t = 1}^T \sum_{i = 1}^{N_{t}} \ln^{2} \widehat{\varepsilon}_{i,t} = 0,
\end{equation}
which implies%
  \footnote{
  Alternative estimators are possible.
  For example, the zero expected score condition $E \left( \ln \varepsilon_{i,t} | \info{F}{i-1, t}\right) = -V/2$ justifies the Method of Moments (MM) estimator 
  \begin{equation}
    \label{eqn:VMM-1}
    \widehat{V} = - 2 N^{-1} \sum_{t = 1}^T \sum_{i = 1}^{N_{t}} \ln \widehat{\varepsilon}_{i,t},
\end{equation}
which is non-negative because of the Jensen's inequality ($E \left( \varepsilon_{i,t} | \info{F}{i-1, t} \right) = 1 \Rightarrow E \left( \ln \varepsilon_{i,t} | \info{F}{i-1, t} \right) \leq \ln E \left( \varepsilon_{i,t} | \info{F}{i-1, t} \right) = \ln 1 = 0$).

Another possibility is to refer again to the first order condition (\ref{eqn:VML}) but replacing the $V^{2} / 4$ addend by the squared average of the $\ln \widehat{\varepsilon}_{i,t}$'s (justified by the zero expected score condition, again).
This leads to estimate $V$ by the sample variance of the $\ln \widehat{\varepsilon}_{i,t}$'s.} 

\begin{equation}
  \label{eqn:VML-2}
  \widehat{V}_{ML} = 2 \left( \sqrt{ N^{-1} \sum_{t = 1}^T \sum_{i = 1}^{N_{t}} \ln^{2} \widehat{\varepsilon}_{i,t}  + 1 } - 1 \right).%
\end{equation}

Because of (\ref{eqn:Ortho:V}), the asymptotic variance matrix of $\widehat{\bm{\theta}}_{ML}$ and $\widehat{V}_{ML}$ depends on their joint behavior.
Assuming the correct specification of $f_{\varepsilon}(\varepsilon_{i,t} | \info{F}{i-1,t})$, the joint \textit{Hessian} matrix is given by
\begin{equation}
  \label{eqn:}
  -V^{-1}
  \left(
  \begin{array}{cc}
    \bm{A}  & - \bm{a}^{\prime} \\
    -\bm{a} & \displaystyle \frac{V+2}{4 V}  
  \end{array}
  \right)
\end{equation}
where
\begin{equation*}
  \bm{a} = \lim_{N \rightarrow \infty} \left[ N^{-1} \sum_{t = 1}^{T} \sum_{i = 1}^{N_{t}} E \left( \bm{a}_{i,t} \right) \right].
\end{equation*}
This implies 
\begin{equation*}
  \avar( \widehat{\bm{\theta}}_{ML}) = V \left( \bm{A} - \frac{V}{V+2} \bm{a}\bm{a}^{\prime} \right)^{-1},
\end{equation*}
which can be estimated by
\begin{equation*}
  \widehat{\avar}( \widehat{\bm{\theta}}_{ML}) = 
  \widehat{V} \left( 
  \widehat{\bm{A}} - \frac{\widehat{V}}{\widehat{V}+2} \widehat{\bm{a}} \widehat{\bm{a}}^{\prime} 
  \right)^{-1},
\end{equation*}
where
\begin{equation*}
  \widehat{\bm{a}} = N^{-1} \sum_{t = 1}^{T} \sum_{i = 1}^{N_{t}} \widehat{\bm{a}}_{i,t}.
\end{equation*}

The availability of several different closed form estimators of $V$ (depending on the $\widehat{\varepsilon}_{i,t}$'s) allows for the possibility to build a concentrated log-likelihood by replacing $V$ with the desired $\widehat{V}$ formula: since the concentrated log-likelihood depends only on $\bm{\theta}$, this bypasses the need to alternate between $\bm{\theta}$ and $V$ estimation \citep[e.g. expression (\ref{eqn:VML-2}) as in ][]{Cattivelli:Gallo:2020}. 
A simpler alternative is maybe to resort to the Method of Moments (MM) estimator (\ref{eqn:VMM-1}), which is also in line with the zero expected score requirement in (\ref{eqn:Cond-1}).

\subsection{Generalized Method of Moments Inference} 
\label{sect:GMMInference}

A different way to estimate the model, which does not need an explicit choice of the error term distribution, is to resort to Generalized Method of Moments (GMM). 
Let
\begin{equation} \label{eq:residuals}
   \varepsilon_{i,t} - 1 = \frac{x_{i,t}}{\tau_{i,t} \xi_{i,t}} - 1.
\end{equation}
Under model assumptions, $\varepsilon_{i,t} - 1$ is a conditionally homoskedastic martingale difference, with conditional expectation zero and conditional variance $\sigma^2$.
Following \citet[][Section~9.2.2.2]{Brownlees:Cipollini:Gallo:2012}, we get that the \emph{efficient} GMM estimators of $\bm{\theta}$, say $\widehat{\bm{\theta}}_{GMM}$, solves the criterion equation (\ref{eqn:Score:Equation:Gamma}) and has the asymptotic variance matrix given in (\ref{eqn:Avar:Sandwich}), i.e., the same properties of $\widehat{\bm{\theta}}_{ML}$ assuming Gamma distributed errors.

In the spirit of a semiparametric approach, a straightforward estimator for  $\sigma^2$ is
\begin{equation*}
  \label{eq:GMMofSigma}
  \widehat{\sigma}^2 = N^{-1} \sum_{t = 1}^{T} \sum_{i = 1}^{N_{t}} \left( \widehat{\varepsilon}_{i,t} - 1 \right)^2
\end{equation*}
where $\widehat{\varepsilon}_{i,t}$ represents here $\varepsilon_{i,t}$ evaluated at $\widehat{\bm{\theta}}_{GMM}$. 
Note that this estimator does not suffer from the presence of zeros in the data.

\section{Empirical Analysis}
\label{sect:Empirical}

Volatility, our main object of interest, is expressed as the square root of the realized kernel variance \citep{BarndorffNielsen:Hansen:Lunde:Shephard:2008, BarndorffNielsen:Hansen:Lunde:Shephard:2009} converted in percentage annualized terms: for the sake of comparison, given that the realized volatility refers to the open--to--close period, we will estimate the GARCH models also in reference to such period. Data on the S\&P 500, FTSE 100, NASDAQ and Hang Seng indices have been collected from the realized library of the Oxford-Man Institute \citep{Heber2009}, which allows us to derive open--to--close returns and their sign. The MIDAS--related macroeconomic variable is the US Industrial Production ($IP_t$), observed monthly and taken from the Federal Reserve Economic Data database. The variable $IPc_t$ is used in month-to-month percentage change \citep[as in][]{Conrad:Loch:2015}. The period under consideration for all the variables is from 2 January 2001 to 15 May 2020. For reference purposes, some summary statistics (minimum, maximum, mean, standard deviation, skewness and kurtosis) for all variables considered are in Table \ref{tab:sum_stat_1}. 

\begin{table}[t]
	\centering
		\caption{Summary statistics}
	\vspace{-0.35cm}
	\label{tab:sum_stat_1}
 \begin{adjustbox}{max height=0.7\textheight , max width=\textwidth}
  \begin{threeparttable}
\begin{tabular}{l  .......}
\toprule
 &     \mc{Obs.} &          \mc{Min.} &         \mc{Max.} &      \mc{Mean} &     \mc{SD} &       \mc{Skew.} & \mc{Kurt.}	\\
\midrule
 \multicolumn{8}{l}{\emph{Daily data}}\\
 S\&P 500 log-returns  & \mc{4859} &  -148.444 & 162.241 & 0.153 & 17.822 & -0.206 & 8.969  \\
 S\&P 500 Realized Kern. Vol. &\mc{4859} & 1.506 & 113.455 & 12.820 & 10.045 & 3.305 & 17.337 \\

 \addlinespace
 \hdashline
 \addlinespace
  FTSE 100 log-returns  & \mc{4884} & -163.500 & 149.018 & -0.118 & 18.405 & -0.404 & 7.921 \\
  FTSE 100  Realized Kern. Vol. &\mc{4884} &   2.231 & 149.437 & 14.574 & 10.467 & 3.938 & 28.319\\

  \addlinespace
 \hdashline
 \addlinespace
NASDAQ log-returns  & \mc{4856} &  -114.995 & 110.062 & 0.050 & 19.038 & -0.221 & 4.218 \\
NASDAQ Realized Kern. Vol. &\mc{4856} & 1.648 & 116.780 & 14.628 & 10.053 & 2.786 & 13.090 \\
 \addlinespace
 \hdashline
 \addlinespace
Hang Seng log-returns  & \mc{4742} &  -184.402 & 192.959 & -0.564 & 15.979 & 0.246 & 12.821\\
Hang Seng Realized Kern. Vol. &\mc{4742} & 2.137 & 130.854 & 13.039 & 8.018 & 3.800 & 28.226 \\

 \addlinespace
 \hdashline
 \addlinespace
 \multicolumn{8}{l}{\emph{Monthly data}}\\
$IPc_t$ & \mc{233} & -346.410 & 5.255 & -1.508 & 22.976 & -14.538 & 214.982 \\
\bottomrule
\addlinespace
 \end{tabular}
 \begin{tablenotes}[flushleft]
   \setlength\labelsep{0pt}
   \footnotesize
\item \textbf{Notes}:  The table reports the number of observations (Obs.), the minimum (Min.) and maximum (Max.), the mean, standard deviation (SD), Skewness (Skew.) and excess Kurtosis (Kurt.). The sample period is 2 January 2001 - 15 May 2020. The daily variables are the open-to-close log-returns and realized kernel volatility, both expressed in annualized percentage. The monthly variable is the US Industrial Production ($IP_t$), and is expressed as the annualized month-to-month percentage change ($IPc_t$), that is $12^{0.5}\cdot 100\cdot((IP_t/IP_{t-1})-1)$. 
\end{tablenotes}
\end{threeparttable}
\end{adjustbox}
\end{table}

\begin{figure}[t]
	\centering
	\caption{Annualized daily log-returns and realized kernel volatility} 
	\label{fig:log_ret_rvol}
	\vspace{-0.35cm}
	\begin{subfigure}[b]{0.48\textwidth}
		\centering
		\includegraphics[width=1\linewidth]{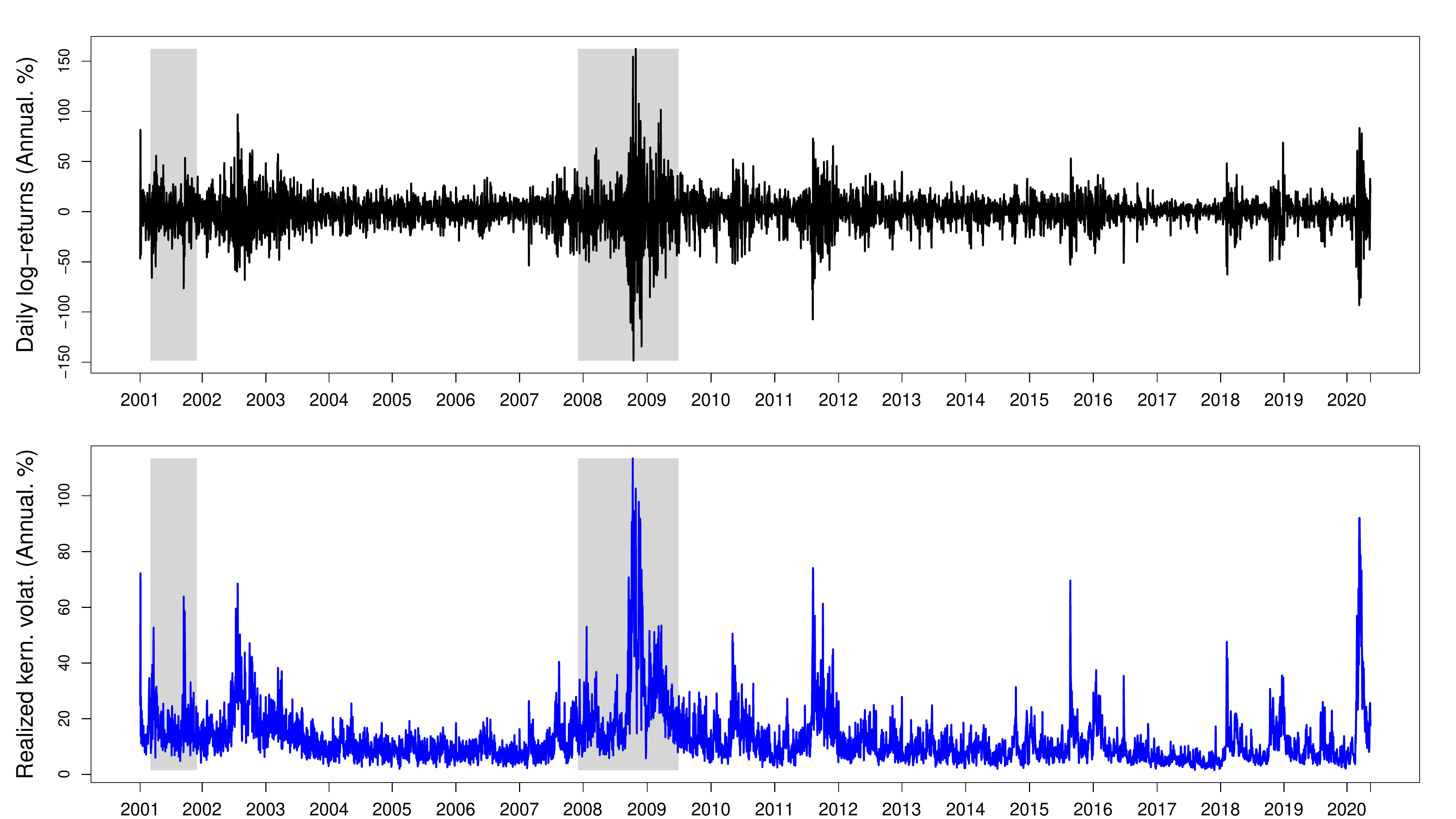}
		\caption{S\&P 500 \label{fig:sp500_ret_vol}}
	\end{subfigure}%
	%\hfill
	\begin{subfigure}[b]{0.48\textwidth}
		\centering
		\includegraphics[width=1\linewidth]{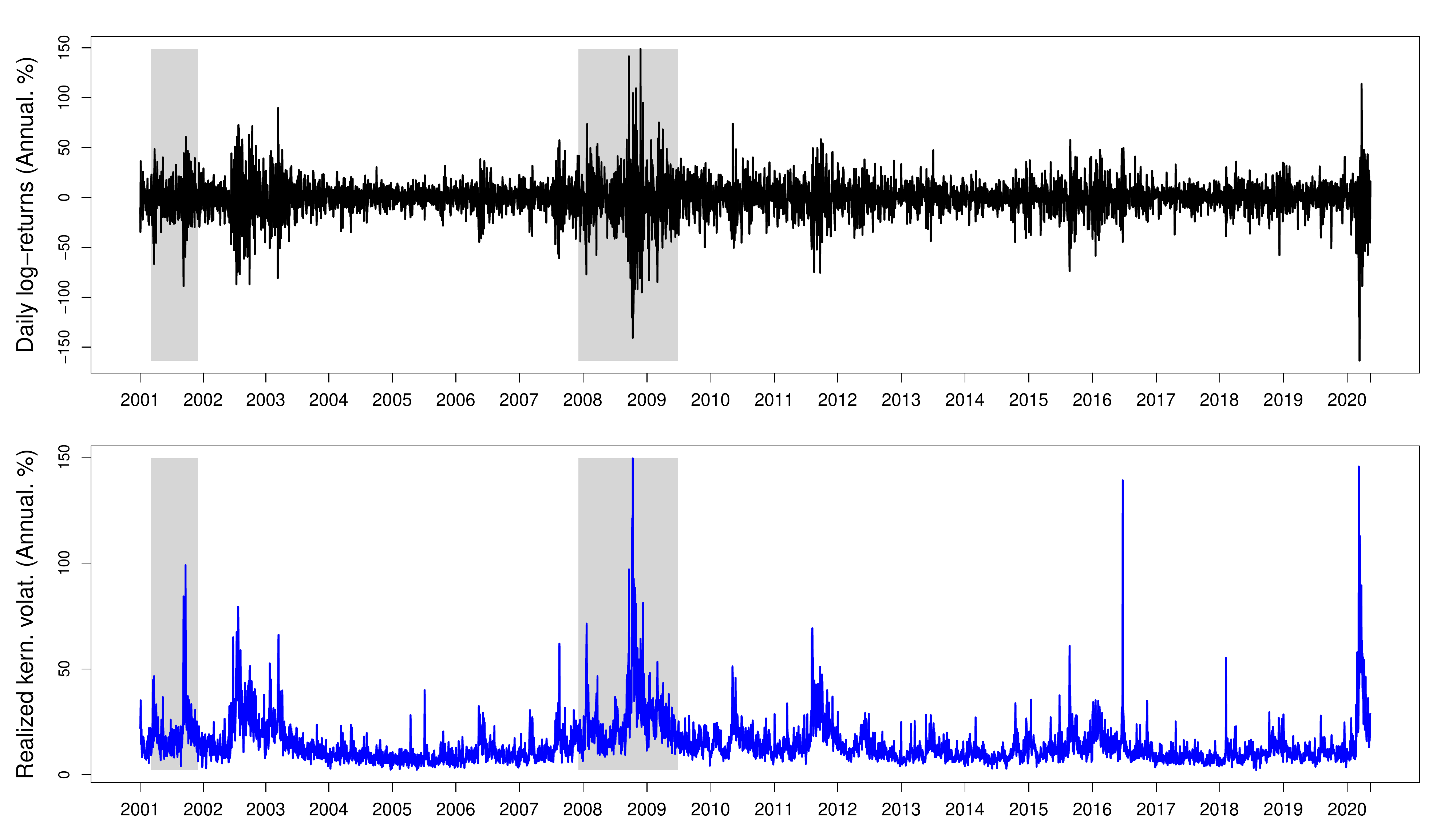}
		\caption{FTSE 100 \label{fig:ftse_100_ret_vol}}
	\end{subfigure}
	\vskip\baselineskip
	\begin{subfigure}[b]{0.48\textwidth}
		\centering
		\includegraphics[width=1\linewidth]{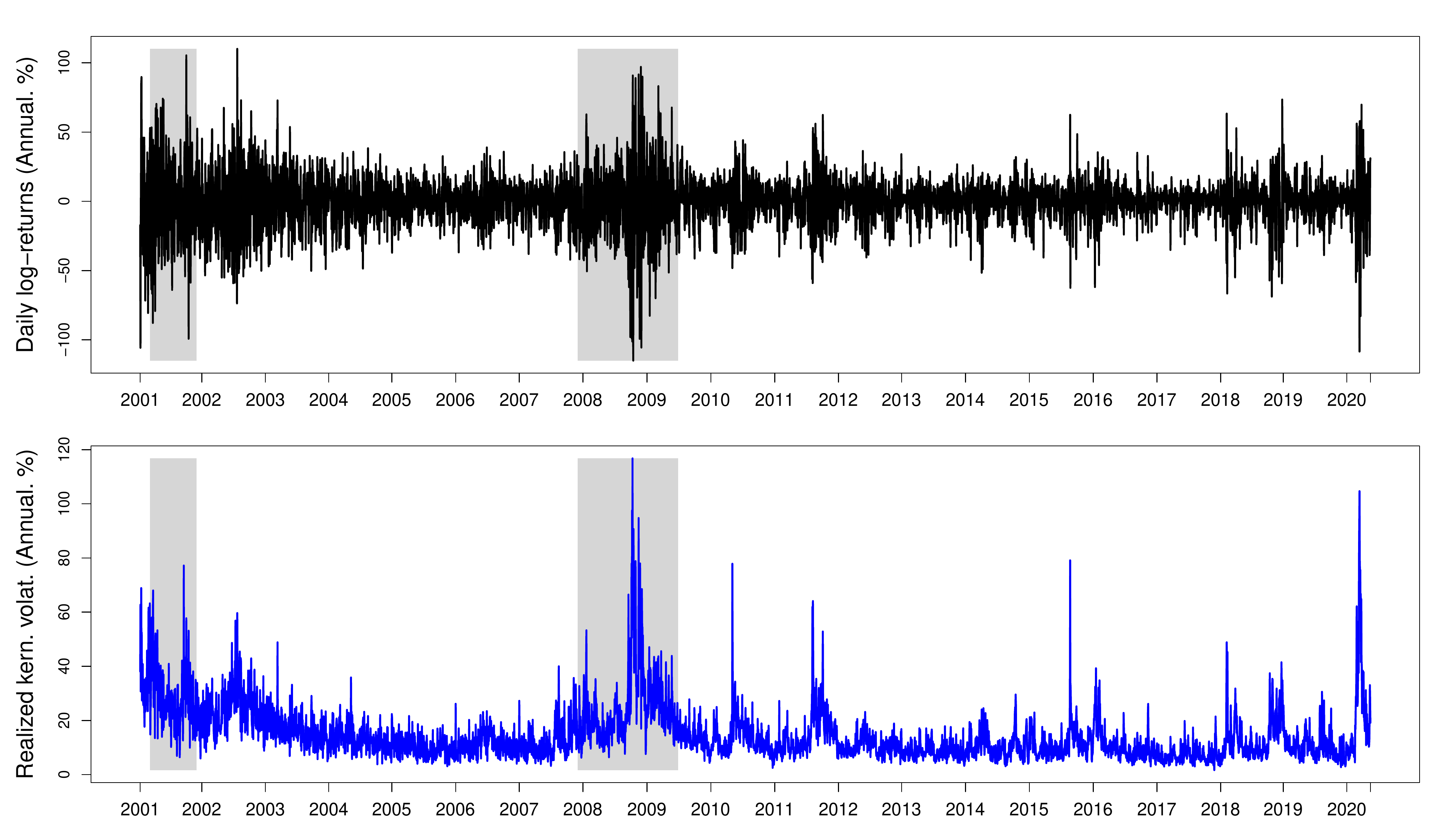}
		\caption{NASDAQ \label{fig:nasdaq_ret_vol}}
	\end{subfigure}
	%\hfill
	\begin{subfigure}[b]{0.48\textwidth}
		\centering
		\includegraphics[width=1\linewidth]{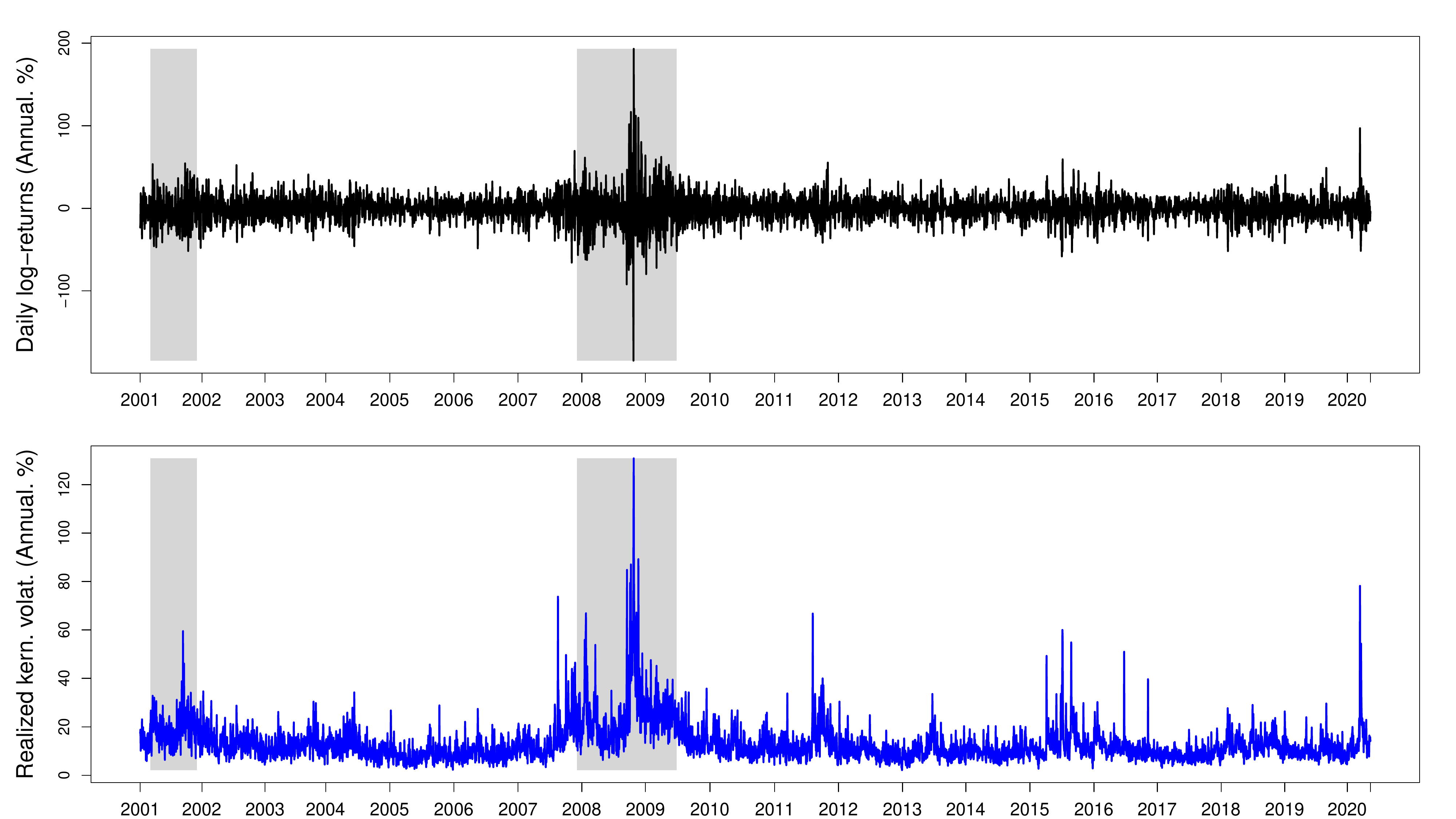}
		\caption{Hang Seng \label{fig:hang_seng_ret_vol}}
	\end{subfigure}%
		\vspace{-0.1cm}
	\begin{minipage}[t]{1\textwidth}
		\textbf{Notes:} Plots of open-to-close log-returns (top panels, black lines) and realized kernel volatilities (bottom panels, blue lines). Shaded areas represent US recession periods (NBER dating).
	\end{minipage}
	
\end{figure}

Figure \ref{fig:log_ret_rvol} depicts the open-to-close log-returns (top panels, black lines) and realized kernel volatilities (bottom panels, blue lines) for the four indices considered over the full sample. We superimposed the US recession periods dated by the NBER in 2001 and then 2008-09, as a reference to periods of slowdown in economic activity (and hence a downturn in industrial production). Although the scales are different, there are features in the dynamics of the series which are common to all four indices, notably the explosion of volatility around the Lehman Brothers demise in September 2008, and other episodes which are more idiosyncratic, although the surge in volatility at the end of 2002 is common to the US and UK indices, and the one in 2015 seems to have affected more the US markets and Hong Kong. 

We include in the set of competing models those having the realized volatility as the dependent variable, namely the multiplicative class (the \AMEM\ plus the two proposed specifications \MEMMIDAS\ and \MEMc) and the asymmetric version of the HAR model (\AHAR), on the one side; and the GARCH class for the conditional variance of open--to--close returns, namely, \GJR, \GM, and the \DAGM, on the other. To the latter, we add the \RGARCH, which is still specified as a GARCH, but makes use of realized variance in its specification. All the functional forms are described in Table \ref{tab:models_eq}. 

\begin{table}[t]
  \centering
  \caption{Model specifications \label{tab:models_eq}}
  \vspace{-0.35cm}
  \begin{adjustbox}{max height=1.0\textheight , max width=1.0\textwidth}
    \begin{threeparttable}
      \begin{tabular}{l c c}
        \toprule  
        Model      &   Functional form     & Err. Distr.\\
        \midrule                                                                                                          
        & $rvol_{i,t}|\mathcal{F}_{i-1,t}  =\mu_{i,t}  \epsilon_{i,t}$ & $\epsilon_{i,t}\overset{i.i.d}{\sim} D^{+}\left(1, \sigma^2\right)$\\          \AMEM  &     $\mu_{i,t}= \alpha_0 + (\alpha_{1} + \gamma_{1} \mathbbm{1}_{\left(r_{i-1,t}  < 0 \right)}) rvol_{i-1,t} +   \beta_{1} \mu_{i-1,t}$& \\
        & $\alpha_0=\left(1-\alpha_{1} - \beta_{1} - \gamma_{1} / 2 \right)\mu$, with $\mu=E\left[ rvol_{i,t} \right]$\\        
    \addlinespace
        \hdashline
        \addlinespace      
      & $rvol_{i,t}|\mathcal{F}_{i-1,t}  =\tau_{i,t} \xi_{i,t} \varepsilon_{i,t}$ & $\epsilon_{i,t}\overset{i.i.d}{\sim} D^{+}\left(1, \sigma^2\right)$\\          \MEMc  &     $\xi_{i,t} = \left(1 - \alpha_{1} - \gamma_{1} / 2 - \beta_{1} \right) + \alpha_{1} x^{(\xi)}_{i-1,t} + \gamma_{1} x^{(\xi-)}_{i-1,t} + \beta_{1} \xi_{i-1,t}$, with $x^{(\xi)}_{i,t}  \equiv \frac{rvol_{i,t}}{\tau_{i,t}} \text{ and }
  x^{(\xi-)}_{i,t} \equiv x^{(\xi)}_{i,t} \mathbbm{1}_{\left(r_{i,t}  < 0 \right)}$& \\
        &  $\tau_{i,t} = \omega^{(\tau)} + \alpha_{1}^{(\tau)} x^{(\tau)}_{i-1,t} + \gamma_{1}^{(\tau)} x^{(\tau-)}_{i-1,t} + \beta_{1}^{(\tau)} \tau_{i-1,t}$, with $x^{(\tau)}_{i,t}  \equiv \frac{rvol_{i,t}}{\xi_{i,t}} \text{ and }
    x^{(\tau-)}_{i,t} \equiv x^{(\tau)}_{i,t} \mathbbm{1}_{\left(r_{i,t}  < 0 \right)}$\\
        \addlinespace
        \hdashline
        \addlinespace

        & $rvol_{i,t}|\mathcal{F}_{i-1,t}  = \tau_t \xi_{i,t}  \epsilon_{i,t}$ & $\epsilon_{i,t}\overset{i.i.d}{\sim} D^{+}\left(1, \sigma^2\right)$\\    
        \MEMMIDAS   &      $\xi_{i,t}=(1-\alpha_{1}-\beta_{1}-\gamma_{1}/2) + \left(\alpha_{1} +  \gamma_{1} \cdot \mathbbm{1}_{\left(r_{i-1,t}  < 0 \right)}\right) \frac{rvol_{i-1,t}}{\tau_t} + \beta_{1} \xi_{i-1,t}$& \\
        & $\tau_t=\exp\left\lbrace m+ \zeta \sum_{k=1}^K \delta_k(\omega) X_{t-k}\right\rbrace$\\

        \addlinespace
        \hdashline
        \addlinespace
        \AHAR &$rvol_{i,t}= c+ (\beta_{1} + \gamma_{1}  \mathbbm{1}_{\left(r_{i-1,t}  < 0 \right)}) \overline{rvol}_{i-1,t}+\beta_5 \overline{rvol}_{(i-2):(i-5),t}+\beta_{22} \overline{rvol}_{(i-6):(i-22),t} + u_{i,t}$ & $u_{i,t} \overset{i.i.d}{\sim} N\left(0, \sigma_{u}^2\right)$ \\
        \addlinespace
        \hdashline
        \addlinespace
        \multirow{2}{*}{\GJR}              & $r_{i,t}|\mathcal{F}_{i-1}  = \sqrt{h_{i,t}} \eta_{i,t}$ & $\eta_{i,t}\overset{i.i.d}{\sim} N\left(0, 1\right)$\\  
           &      $h_{i,t} = const + \left(\alpha_{1} + \gamma_{1} \mathbbm{1}_{\left( r_{i-1, t} < 0 \right)} \right) r_{i-1, t}^2 + \beta_{1} h_{i-1, t}$& \\
          \addlinespace
        \hdashline
        \addlinespace  
        & $r_{i,t}|\mathcal{F}_{i-1,t}  = \sqrt{\tau_t \times \xi_{i,t} } \eta_{i,t}$ & $\eta_{i,t}\overset{i.i.d}{\sim} N\left(0, 1\right)$\\    
        \GM   &      $\xi_{i,t}=(1-\alpha_{1}-\beta_{1}-\gamma_{1}/2) + \left(\alpha_{1} +  \gamma_{1} \cdot \mathbbm{1}_{\left(r_{i-1,t}  < 0 \right)}\right) \frac{r_{i-1,t}^2}{\tau_t} + \beta_{1} \xi_{i-1,t}$& \\
        & $\tau_t=\exp\left\lbrace m + \zeta \sum_{k=1}^K \delta_k(\omega) X_{t-k}\right\rbrace$\\
        \addlinespace
        \hdashline
        \addlinespace  
        & $r_{i,t}|\mathcal{F}_{i-1,t}  = \sqrt{\tau_t \times \xi_{i,t} } \eta_{i,t}$ & $\eta_{i,t}\overset{i.i.d}{\sim} N\left(0, 1\right)$\\  
        \DAGM   &      $\xi_{i,t}=(1-\alpha_{1}-\beta_{1}-\gamma_{1}/2) + \left(\alpha_{1} +  \gamma_{1} \cdot \mathbbm{1}_{\left(r_{i-1,t}  < 0 \right)}\right) \frac{r_{i-1,t}^2}{\tau_t} + \beta_{1} \xi_{i-1,t}$& \\
        & $\tau_t=\exp\left\lbrace m +
        \zeta^{+}  \sum_{k=1}^K \delta_k(\omega)^{+}  X_{t-k} \mathbbm{1}_{\left( X_{t-k} \geq 0 \right)} +  \zeta^{-}  \sum_{k=1}^K \delta_k(\omega)^{-} X_{t-k} \mathbbm{1}_{\left( X_{t-k} < 0 \right)} \right\rbrace$\\
        \addlinespace
        \hdashline
        \addlinespace
  
  \multirow{2}{*}{\RGARCH}      & $r_{i, t}|\mathcal{F}_{i-1,t}  = \sqrt{h_{i,t}} \eta_{i, t}$ & $\eta_{i, t}\overset{i.i.d}{\sim} N\left(0, 1\right)$\\  
        &      $\log(h_{i, t}) = const + \beta_{1} \log(h_{i-1, t}) + \alpha_{1} \log(rvol_{i-1, t})$& \\
%        & $rvol_i=\nu + \varphi h_i + \pi \eta_{i} + \pi \left(\eta_{i}^2-1\right) +\sigma_z z_i$ &$z_i\overset{i.i.d}{\sim} N\left(0, 1\right)$\\
        \bottomrule
      \end{tabular}
      \begin{tablenotes}[flushleft]
        \setlength\labelsep{0pt}
        \footnotesize
        \item \textbf{Notes}: The table reports the functional forms for the Asymmetric MEM (\AMEM), \MEMMIDAS, \MEMc, 
Asymmetric HAR (\AHAR),  \GJR, GARCH--MIDAS (\GM),  Double Asymmetric GARCH--MIDAS (\DAGM),  and Realized GARCH (\RGARCH) specifications.
      \end{tablenotes}
    \end{threeparttable}
  \end{adjustbox}
\end{table}

The testing ground for the models includes two different robust loss functions (LFs, \cite{Patton:2011}): QLIKE and MSE. All LFs have the realized kernel volatility as their target, and the GARCH models variance forecasts are modified to match that target. The evaluation makes use of the Model Confidence Set \citep[MCS,][]{Hansen:Lunde:Nason:2011}, and the test statistic used in the MCS procedure is the semi-quadratic $T_{SQ}$, as recently done by  \cite{Cipollini:Gallo:Otranto:2020}, for instance.

\subsection{In--sample analysis}

The first in--sample period spans from January 2001 to December 2012. Tables from \ref{tab:in_sample_est_sp500} to \ref{tab:in_sample_est_hang_seng} report the estimated coefficients for each model, some residual diagnostics and the MCS inclusion according to the two LFs. In terms of diagnostics, we consider the Ljung-Box \citep{Ljung:Box:1978}, applied on standardized residuals (squared standardized residuals for the GARCH-based models) at different lags.
Overall, considering higher lags, the tests for the two proposed specifications signal an absence of clustering in the residuals (except for the NASDAQ index), contrary to what happens for many  of the other competing specifications. As regards to the inclusion in the MCS, we can notice that MEM--based specifications have a better performance than all the other models. Interestingly, the proposed \MEMc\ model is always included in the set of the superior models, independently of the LF adopted. In the case of the FTSE 100, the \MEMc\ model is the only specification belonging to the MCS.

\begin{table}[htbp]
\centering
\caption{In-sample comparison. S\&P 500  \label{tab:in_sample_est_sp500}}
\vspace{-0.35cm}
 \begin{adjustbox}{max height=0.7\textheight , max width=\textwidth}
  \begin{threeparttable}
\begin{tabular}{l ........}
\toprule  
\textbf{}               	&	  \mc{\AMEM}         &	       \mc{\MEMc}   	&	   \mc{\MEMMIDAS}          	&	       \mc{\AHAR}            	     	&	  \mc{\GJR}   &       \mc{\GM}   & 	         \mc{\DAGM}       &	         \mc{\RGARCH}       \\
\midrule                                                                                                          
$const$ &                 0.296 &          &          &  1.016^{**} & 4.056^{***}  &          &           &  0.25^{***} \\
&                           &          &          &     (0.466) &     (1.011)  &          &           &     (0.079) \\
$\alpha_1$ &          0.1^{***} & -0.112^{***} &  0.092^{***} &         &       0^{ }  &   0.001^{ }  &    0.001^{ }  & 0.389^{***} \\
 &                      (0.008) &       (0.02) &       (0.02) &         &     (0.013)  &     (0.019)  &      (0.014)  &     (0.038) \\
$\beta_1$ &         0.823^{***} &  0.518^{***} &   0.82^{***} &  0.17^{***} & 0.914^{***}  & 0.912^{***}  &  0.914^{***}  & 0.777^{***} \\
  &                     (0.009) &        (0.1) &      (0.022) &     (0.041) &     (0.017)  &     (0.016)  &      (0.016)  &     (0.022) \\
$\beta_5$ &                 &          &          & 0.524^{***} &          &          &           &         \\
   &                        &          &          &     (0.061) &          &          &           &         \\
$\beta_{22}$ &              &          &          &  0.15^{***} &          &          &           &         \\
     &                      &          &          &     (0.055) &          &          &           &         \\
$\gamma_1$ &         0.113^{***} &  0.106^{***} &   0.12^{***} & 0.172^{***} & 0.142^{***}  & 0.144^{***}  &  0.145^{***}  &         \\
&                       (0.006) &      (0.014) &      (0.015) &     (0.022) &      (0.02)  &      (0.02)  &       (0.02)  &         \\
$m$ &                       &          &    0.009^{ } &         &          &  5.73^{***}  &   5.63^{***}  &         \\
&                           &          &      (0.076) &         &          &     (0.346)  &      (0.693)  &         \\
$\zeta$ &                    &          & -0.164^{***} &         &          &  -0.225^{ }  &           &         \\
&                           &          &      (0.058) &         &          &     (0.182)  &           &         \\
$\omega_2$ &                 &          &    4.11^{**} &         &          &  1.445^{**}  &           &         \\
&                           &          &      (1.797) &         &          &     (0.708)  &           &         \\
$\zeta^+$ &                  &          &          &         &          &          & -1.024^{***}  &         \\
 &                          &          &          &         &          &          &      (0.381)  &         \\
$\omega_2^+$ &               &          &          &         &          &          &  1.424^{***}  &         \\
  &                         &          &          &         &          &          &      (0.344)  &         \\
$\zeta^-$ &                  &          &          &         &          &          &  -1.13^{***}  &         \\
  &                         &          &          &         &          &          &       (0.38)  &         \\
$\omega_2^-$ &               &          &          &         &          &          &  1.001^{***}  &         \\
   &                        &          &          &         &          &          &      (0.364)  &         \\
$\omega^{(\tau)}$ &            &  0.256^{***} &          &         &          &          &           &         \\
    &                       &      (0.043) &          &         &          &          &           &         \\
$\alpha_1^{(\tau)}$ &         &  0.123^{***} &          &         &          &          &           &         \\
    &                       &      (0.015) &          &         &          &          &           &         \\
$\beta_1^{(\tau)}$ &          &  0.819^{***} &          &         &          &          &           &         \\
   &                        &      (0.014) &          &         &          &          &           &         \\
$\gamma_1^{(\tau)}$ &          &   0.08^{***} &          &         &          &          &           &         \\
     &                      &      (0.011) &          &         &          &          &           &         \\

\addlinespace																		
\hdashline																		
\addlinespace 													LB$_5$ &                  0.001 &        0.007 &        0.001 &           0.000 &       0.001  &       0.002  &        0.001  &       0.006 \\
LB$_{10}$ &               0.012 &        0.045 &        0.017 &           0.000 &       0.007  &       0.016  &        0.006  &       0.009 \\
LB$_{20}$ &               0.087 &        0.113 &        0.134 &           0.000 &       0.084  &       0.136  &        0.071  &       0.102 \\

QLIKE&   0.069 &\cellcolor{gray!75}0.067 & 0.068 & 0.071 & 0.084 & 0.084 & 0.085 & 0.087 \\

MSE&  0.16 & \cellcolor{gray!75}0.157 & \cellcolor{gray!75}0.157 & 0.164 & 0.199 & 0.203 & 0.206 & 0.199 \\ 

     \bottomrule
\end{tabular}
  \begin{tablenotes}[flushleft]
   \setlength\labelsep{0pt}
   \footnotesize
\item \textbf{Notes}: The table reports the estimated coefficients of the models in column. $^{*}$, $^{**}$ and $^{***}$ represent the significance at levels $10\%, 5\%, 1\%$, respectively, associated to QML standard errors. The reported constant for the \AMEM\ model refers to $\alpha_{0}$ parameter in Table~\ref{tab:models_eq}. For ease of notation, the parameter $\alpha_1$ referred to the \RGARCH\ corresponds to the parameter labelled as $\gamma$ in \cite{Hansen:Huang:Shek:2012}. Moreover, the estimated parameters of the measurement equation of this latter model are not reported for space constraints. LB$_{l}$  represents the p-values of the Ljung-Box \citep{Ljung:Box:1978} test at $l$ lag, applied on standardized residuals (squared for GARCH models).  Last two rows report the averages of the QLIKE and MSE loss functions. The chosen volatility proxy is the realized kernel. Shades of gray denote inclusion in the MCS at significance level $\alpha=0.25$.
\item Sample period: January 2001 - December 2012. Daily observations: 3008. Macro-economic variable for the MIDAS model: $IPc_t$. Number of lagged macro-economic variable realizations: $K=36$.
\end{tablenotes}
\end{threeparttable}
\end{adjustbox}
\end{table}

\begin{table}[htbp]
\centering
\caption{In-sample comparison. FTSE 100 \label{tab:in_sample_est_ftse_100}}
\vspace{-0.35cm}
 \begin{adjustbox}{max height=0.7\textheight , max width=\textwidth}
  \begin{threeparttable}
\begin{tabular}{l ........}
\toprule  
\textbf{}               	&	  \mc{\AMEM}         &	       \mc{\MEMc}   	&	   \mc{\MEMMIDAS}          	&	       \mc{\AHAR}            	     	&	  \mc{\GJR}   &       \mc{\GM}   & 	         \mc{\DAGM}       &	         \mc{\RGARCH}       \\
\midrule                                                                                                          
$const$ &                 0.311 &         &          & 1.097^{***} &  4.13^{***}  &           &          &  0.12^{***} \\
&                           &         &          &     (0.422) &     (0.843)  &           &          &     (0.041) \\
$\alpha_1$ &        0.149^{***} &   0.019^{ } &  0.136^{***} &         &       0^{ }  &    0.001^{ }  &   0.001^{ }  & 0.478^{***} \\
 &                       (0.01) &     (0.023) &       (0.01) &         &      (0.01)  &      (0.018)  &     (0.016)  &     (0.032) \\
$\beta_1$ &         0.779^{***} & 0.649^{***} &  0.781^{***} & 0.252^{***} & 0.901^{***}  &  0.891^{***}  & 0.899^{***}  & 0.751^{***} \\
  &                     (0.011) &     (0.082) &      (0.011) &     (0.039) &     (0.012)  &      (0.017)  &     (0.015)  &     (0.017) \\
$\beta_5$ &                 &         &          & 0.433^{***} &          &           &          &         \\
   &                        &         &          &     (0.056) &          &           &          &         \\
$\beta_{22}$ &              &         &          &  0.18^{***} &          &           &          &         \\
     &                      &         &          &     (0.041) &          &           &          &         \\
$\gamma_1$ &         0.105^{***} & 0.069^{***} &  0.109^{***} & 0.127^{***} & 0.169^{***}  &  0.177^{***}  & 0.171^{***}  &         \\
&                       (0.006) &     (0.014) &      (0.006) &     (0.024) &     (0.024)  &      (0.024)  &     (0.024)  &         \\
$m$ &                       &         &  -0.075^{**} &         &          &  5.694^{***}  & 4.794^{***}  &         \\
&                           &         &      (0.035) &         &          &      (0.208)  &     (1.225)  &         \\
$\zeta$ &                    &         & -0.188^{***} &         &          & -0.444^{***}  &          &         \\
&                           &         &      (0.031) &         &          &      (0.094)  &          &         \\
$\omega_2$ &                 &         &   1.77^{***} &         &          &   1.45^{***}  &          &         \\
&                           &         &      (0.193) &         &          &      (0.398)  &          &         \\
$\zeta^+$ &                  &         &          &         &          &           &   0.658^{ }  &         \\
 &                          &         &          &         &          &           &     (0.808)  &         \\
$\omega_2^+$ &               &         &          &         &          &           &  1.001^{**}  &         \\
  &                         &         &          &         &          &           &     (0.474)  &         \\
$\zeta^-$ &                  &         &          &         &          &           &  -0.119^{ }  &         \\
  &                         &         &          &         &          &           &     (0.334)  &         \\
$\omega_2^-$ &               &         &          &         &          &           & 4.713^{***}  &         \\
   &                        &         &          &         &          &           &     (0.485)  &         \\
$\omega^{(\tau)}$ &            & 0.185^{***} &          &         &          &           &          &         \\
    &                       &      (0.04) &          &         &          &           &          &         \\
$\alpha_1^{(\tau)}$ &         & 0.104^{***} &          &         &          &           &          &         \\
    &                       &     (0.017) &          &         &          &           &          &         \\
$\beta_1^{(\tau)}$ &          &  0.85^{***} &          &         &          &           &          &         \\
   &                        &     (0.017) &          &         &          &           &          &         \\
$\gamma_1^{(\tau)}$ &          & 0.067^{***} &          &         &          &           &          &         \\
     &                      &     (0.012) &          &         &          &           &          &         \\

\addlinespace																		
\hdashline																		
\addlinespace 																		
LB$_5$ &                  0.008 &       0.102 &        0.009 &           0.000 &       0.106  &        0.061  &         0.1  &       0.498 \\
LB$_{10}$ &               0.065 &       0.298 &        0.081 &           0.000 &       0.148  &        0.147  &       0.163  &       0.485 \\
LB$_{20}$ &               0.157 &       0.579 &        0.215 &           0.000 &       0.027  &        0.019  &       0.033  &       0.146 \\

QLIKE&   0.053 & \cellcolor{gray!75}0.053 & 0.053 & 0.055 & 0.061 & 0.058 & 0.061 & 0.06 \\

MSE&    0.174 & \cellcolor{gray!75}0.17 & 0.173 & 0.18 & 0.194 & 0.188 & 0.196 & 0.197 \\

    \bottomrule
\end{tabular}
  \begin{tablenotes}[flushleft]
   \setlength\labelsep{0pt}
   \footnotesize
\item \textbf{Notes}:  The table reports the estimated coefficients of the models in column. $^{*}$, $^{**}$ and $^{***}$ represent the significance at levels $10\%, 5\%, 1\%$, respectively, associated to QML standard errors. The reported constant for the \AMEM\ model refers to $\alpha_{0}$ parameter in Table~\ref{tab:models_eq}. For ease of notation, the parameter $\alpha_1$ referred to the \RGARCH\ corresponds to the parameter labelled as $\gamma$ in \cite{Hansen:Huang:Shek:2012}. Moreover, the estimated parameters of the measurement equation of this latter model are not reported for space constraints. LB$_{l}$  represents the p-values of the Ljung-Box \citep{Ljung:Box:1978} test at $l$ lag, applied on standardized residuals (squared for GARCH models).  Last two rows report the averages of the QLIKE and MSE loss functions. The chosen volatility proxy is the realized kernel. Shades of gray denote inclusion in the MCS at significance level $\alpha=0.25$.
\item Sample period: January 2001 - December 2012. Daily observations: 3021. Macro-economic variable for the MIDAS model: $IPc_t$. Number of lagged macro-economic variable realizations: $K=36$.
\end{tablenotes}
\end{threeparttable}
\end{adjustbox}
\end{table}

\begin{table}[htbp]
\centering
\caption{In-sample comparison: NASDAQ \label{tab:in_sample_est_nasdaq}}
\vspace{-0.35cm}
 \begin{adjustbox}{max height=0.7\textheight , max width=\textwidth}
  \begin{threeparttable}
\begin{tabular}{l ........}
\toprule  
\textbf{}               	&	  \mc{\AMEM}         &	       \mc{\MEMc}   	&	   \mc{\MEMMIDAS}          	&	       \mc{\AHAR}            	     	&	  \mc{\GJR}   &       \mc{\GM}   & 	         \mc{\DAGM}       &	         \mc{\RGARCH}       \\
\midrule                                                                                                          
$const$ &                 0.321 &         &         &  0.957^{**} & 3.146^{***}  &           &          & 0.183^{***} \\
&                           &         &         &     (0.374) &     (0.933)  &           &          &     (0.042) \\
$\alpha_1$ &        0.163^{***} &   0.031^{ } & 0.159^{***} &         &   0.016^{*}  &  0.028^{***}  & 0.028^{***}  & 0.419^{***} \\
 &                       (0.01) &     (0.021) &     (0.025) &         &     (0.009)  &       (0.01)  &      (0.01)  &     (0.042) \\
$\beta_1$ &         0.784^{***} & 0.652^{***} & 0.779^{***} & 0.236^{***} &  0.93^{***}  &  0.926^{***}  & 0.928^{***}  & 0.772^{***} \\
  &                     (0.011) &     (0.048) &     (0.027) &     (0.028) &     (0.013)  &       (0.01)  &     (0.009)  &     (0.022) \\
$\beta_5$ &                 &         &         & 0.435^{***} &          &           &          &         \\
   &                        &         &         &     (0.047) &          &           &          &         \\
$\beta_{22}$ &              &         &         & 0.195^{***} &          &           &          &         \\
     &                      &         &         &     (0.048) &          &           &          &         \\
$\gamma_1$ &         0.069^{***} & 0.105^{***} & 0.073^{***} & 0.153^{***} & 0.085^{***}  &  0.089^{***}  & 0.085^{***}  &         \\
&                       (0.006) &     (0.012) &     (0.017) &     (0.019) &     (0.016)  &       (0.02)  &     (0.014)  &         \\
$m$ &                       &         &   0.032^{ } &         &          &  7.381^{***}  &   7.8^{***}  &         \\
&                           &         &     (0.103) &         &          &      (1.224)  &     (0.858)  &         \\
$\zeta$ &                    &         &  -0.128^{ } &         &          & -0.221^{***}  &          &         \\
&                           &         &     (0.079) &         &          &      (0.081)  &          &         \\
$\omega_2$ &                 &         &   5.809^{ } &         &          &  5.667^{***}  &          &         \\
&                           &         &     (4.364) &         &          &      (0.274)  &          &         \\
$\zeta^+$ &                  &         &         &         &          &           &   -0.65^{ }  &         \\
 &                          &         &         &         &          &           &     (0.737)  &         \\
$\omega_2^+$ &               &         &         &         &          &           &   1.921^{ }  &         \\
  &                         &         &         &         &          &           &     (1.924)  &         \\
$\zeta^-$ &                  &         &         &         &          &           &  -0.304^{ }  &         \\
  &                         &         &         &         &          &           &     (0.828)  &         \\
$\omega_2^-$ &               &         &         &         &          &           &   3.484^{ }  &         \\
   &                        &         &         &         &          &           &     (9.729)  &         \\
$\omega^{(\tau)}$ &            & 0.123^{***} &         &         &          &           &          &         \\
    &                       &     (0.037) &         &         &          &           &          &         \\
$\alpha_1^{(\tau)}$ &         & 0.099^{***} &         &         &          &           &          &         \\
    &                       &     (0.015) &         &         &          &           &          &         \\
$\beta_1^{(\tau)}$ &          & 0.881^{***} &         &         &          &           &          &         \\
   &                        &     (0.015) &         &         &          &           &          &         \\
$\gamma_1^{(\tau)}$ &          & 0.025^{***} &         &         &          &           &          &         \\
     &                      &     (0.009) &         &         &          &           &          &         \\

\addlinespace														\hdashline															\addlinespace 								LB$_5$ &                  0.013 &       0.002 &       0.015 &           0.000 &       0.002  &        0.003  &       0.003  &       0.006 \\
LB$_{10}$ &               0.005 &       0.014 &       0.006 &           0.000 &       0.002  &        0.003  &       0.003  &       0.009 \\
LB$_{20}$ &               0.005 &       0.002 &       0.009 &           0.000 &       0.039  &        0.031  &       0.038  &       0.039 \\
QLIKE& 0.053 & \cellcolor{gray!75}0.051 & 0.053 & 0.054 & 0.067 & 0.071 & 0.071 & 0.064 \\ 
MSE & 0.159 & \cellcolor{gray!75}0.153 & 0.157 & 0.159 & 0.196 & 0.229 & 0.226 & 0.194 \\ 
     
  \bottomrule
\end{tabular}
  \begin{tablenotes}[flushleft]
   \setlength\labelsep{0pt}
   \footnotesize
\item \textbf{Notes}:  The table reports the estimated coefficients of the models in column. $^{*}$, $^{**}$ and $^{***}$ represent the significance at levels $10\%, 5\%, 1\%$, respectively, associated to QML standard errors. The reported constant for the \AMEM\ model refers to $\alpha_{0}$ parameter in Table~\ref{tab:models_eq}. For ease of notation, the parameter $\alpha_1$ referred to the \RGARCH\ corresponds to the parameter labelled as $\gamma$ in \cite{Hansen:Huang:Shek:2012}. Moreover, the estimated parameters of the measurement equation of this latter model are not reported for space constraints. LB$_{l}$  represents the p-values of the Ljung-Box \citep{Ljung:Box:1978} test at $l$ lag, applied on standardized residuals (squared for GARCH models).  Last two rows report the averages of the QLIKE and MSE loss functions. The chosen volatility proxy is the realized kernel. Shades of gray denote inclusion in the MCS at significance level $\alpha=0.25$.
\item Sample period: January 2001 - December 2012. Daily observations: 3005. Macro-economic variable for the MIDAS model: $IPc_t$. Number of lagged macro-economic variable realizations: $K=36$.
\end{tablenotes}
\end{threeparttable}
\end{adjustbox}
\end{table}

\begin{table}[htbp]
\centering
\caption{In-sample comparison. Hang Seng \label{tab:in_sample_est_hang_seng}}
\vspace{-0.35cm}
 \begin{adjustbox}{max height=0.7\textheight , max width=\textwidth}
  \begin{threeparttable}
\begin{tabular}{l ........}
\toprule  
\textbf{}               	&	  \mc{\AMEM}         &	       \mc{\MEMc}   	&	   \mc{\MEMMIDAS}          	&	       \mc{\AHAR}            	     	&	  \mc{\GJR}   &       \mc{\GM}   & 	         \mc{\DAGM}       &	         \mc{\RGARCH}       \\
\midrule                                                                                                          
$const$ &                 0.157 &         &          &   0.974^{*} &  2.18^{***}  &          &          & 0.084^{***} \\
&                           &         &          &      (0.53) &     (0.631)  &          &          &      (0.03) \\
$\alpha_1$ &        0.136^{***} &   0.032^{ } &  0.134^{***} &         & 0.035^{***}  & 0.032^{***}  & 0.031^{***}  & 0.303^{***} \\
 &                      (0.009) &     (0.021) &      (0.009) &         &     (0.011)  &     (0.011)  &      (0.01)  &     (0.062) \\
$\beta_1$ &         0.845^{***} & 0.598^{***} &  0.838^{***} & 0.246^{***} & 0.939^{***}  & 0.932^{***}  & 0.942^{***}  & 0.842^{***} \\
  &                      (0.01) &     (0.138) &      (0.011) &     (0.054) &      (0.01)  &     (0.009)  &     (0.008)  &     (0.031) \\
$\beta_5$ &                 &         &          & 0.384^{***} &          &          &          &         \\
   &                        &         &          &     (0.057) &          &          &          &         \\
$\beta_{22}$ &              &         &          & 0.282^{***} &          &          &          &         \\
     &                      &         &          &     (0.046) &          &          &          &         \\
$\gamma_1$ &         0.015^{***} & 0.038^{***} &  0.015^{***} &   0.034^{ } &  0.032^{**}  &  0.034^{**}  &  0.034^{**}  &         \\
&                       (0.006) &     (0.013) &      (0.006) &     (0.023) &     (0.015)  &     (0.016)  &     (0.014)  &         \\
$m$ &                       &         &   -0.031^{ } &         &          & 5.439^{***}  & 6.519^{***}  &         \\
&                           &         &       (0.04) &         &          &     (0.115)  &     (0.389)  &         \\
$\zeta$ &                    &         & -0.155^{***} &         &          & -0.309^{**}  &          &         \\
&                           &         &      (0.028) &         &          &      (0.13)  &          &         \\
$\omega_2$ &                 &         &  4.047^{***} &         &          &    4.33^{ }  &          &         \\
&                           &         &      (0.201) &         &          &     (3.949)  &          &         \\
$\zeta^+$ &                  &         &          &         &          &          &  -0.376^{ }  &         \\
 &                          &         &          &         &          &          &     (0.263)  &         \\
$\omega_2^+$ &               &         &          &         &          &          & 3.006^{***}  &         \\
  &                         &         &          &         &          &          &     (1.131)  &         \\
$\zeta^-$ &                  &         &          &         &          &          &   0.68^{**}  &         \\
  &                         &         &          &         &          &          &     (0.338)  &         \\
$\omega_2^-$ &               &         &          &         &          &          & 1.001^{***}  &         \\
   &                        &         &          &         &          &          &     (0.322)  &         \\
$\omega^{(\tau)}$ &            &  0.11^{***} &          &         &          &          &          &         \\
    &                       &     (0.037) &          &         &          &          &          &         \\
$\alpha_1^{(\tau)}$ &         & 0.101^{***} &          &         &          &          &          &         \\
    &                       &     (0.013) &          &         &          &          &          &         \\
$\beta_1^{(\tau)}$ &          & 0.884^{***} &          &         &          &          &          &         \\
   &                        &     (0.014) &          &         &          &          &          &         \\
$\gamma_1^{(\tau)}$ &          &   0.014^{ } &          &         &          &          &          &         \\
     &                      &     (0.009) &          &         &          &          &          &         \\

\addlinespace																		
\hdashline																		
\addlinespace 																		
LB$_5$ &                  0.078 &       0.436 &        0.107 &       0.001 &       0.907  &       0.897  &       0.899  &       0.162 \\
LB$_{10}$ &                0.04 &       0.732 &        0.068 &       0.001 &        0.26  &       0.426  &       0.219  &       0.117 \\
LB$_{20}$ &               0.059 &       0.398 &        0.114 &           0.000 &        0.28  &       0.594  &        0.23  &       0.167 \\

QLIKE & 0.064 & \cellcolor{gray!75}0.063 & 0.064 & 0.065 & 0.075 & 0.074 & 0.074 & 0.069 \\

MSE& 0.149 & \cellcolor{gray!75}0.146 & \cellcolor{gray!75}0.147 & 0.148 & 0.171 & 0.166 & 0.162 & 0.16 \\ 

 \bottomrule
\end{tabular}
  \begin{tablenotes}[flushleft]
   \setlength\labelsep{0pt}
   \footnotesize
\item \textbf{Notes}: The table reports the estimated coefficients of the models in column. $^{*}$, $^{**}$ and $^{***}$ represent the significance at levels $10\%, 5\%, 1\%$, respectively, associated to QML standard errors. The reported constant for the \AMEM\ model refers to $\alpha_{0}$ parameter in Table~\ref{tab:models_eq}. For ease of notation, the parameter $\alpha_1$ referred to the \RGARCH\ corresponds to the parameter labelled as $\gamma$ in \cite{Hansen:Huang:Shek:2012}. Moreover, the estimated parameters of the measurement equation of this latter model are not reported for space constraints. LB$_{l}$  represents the p-values of the Ljung-Box \citep{Ljung:Box:1978} test at $l$ lag, applied on standardized residuals (squared for GARCH models).  Last two rows report the averages of the QLIKE and MSE loss functions. The chosen volatility proxy is the realized kernel. Shades of gray denote inclusion in the MCS at significance level $\alpha=0.25$.
\item Sample period: January 2001 - December 2012. Daily observations: 2938. Macro-economic variable for the MIDAS model: $IPc_t$. Number of lagged macro-economic variable realizations: $K=36$.
\end{tablenotes}
\end{threeparttable}
\end{adjustbox}
\end{table}

\subsection{A Graphical Appraisal of the \lrc}

\begin{figure}[t]
	\centering
	\caption{Monthly $\tau_{t}$ term: comparison between \MEMc\ and \MEMMIDAS} 
	\label{fig:cmem_mem_midas_tau}
	\vspace{-0.35cm}
\begin{subfigure}[b]{1\textwidth}
		\centering
		\includegraphics[width=1\linewidth]{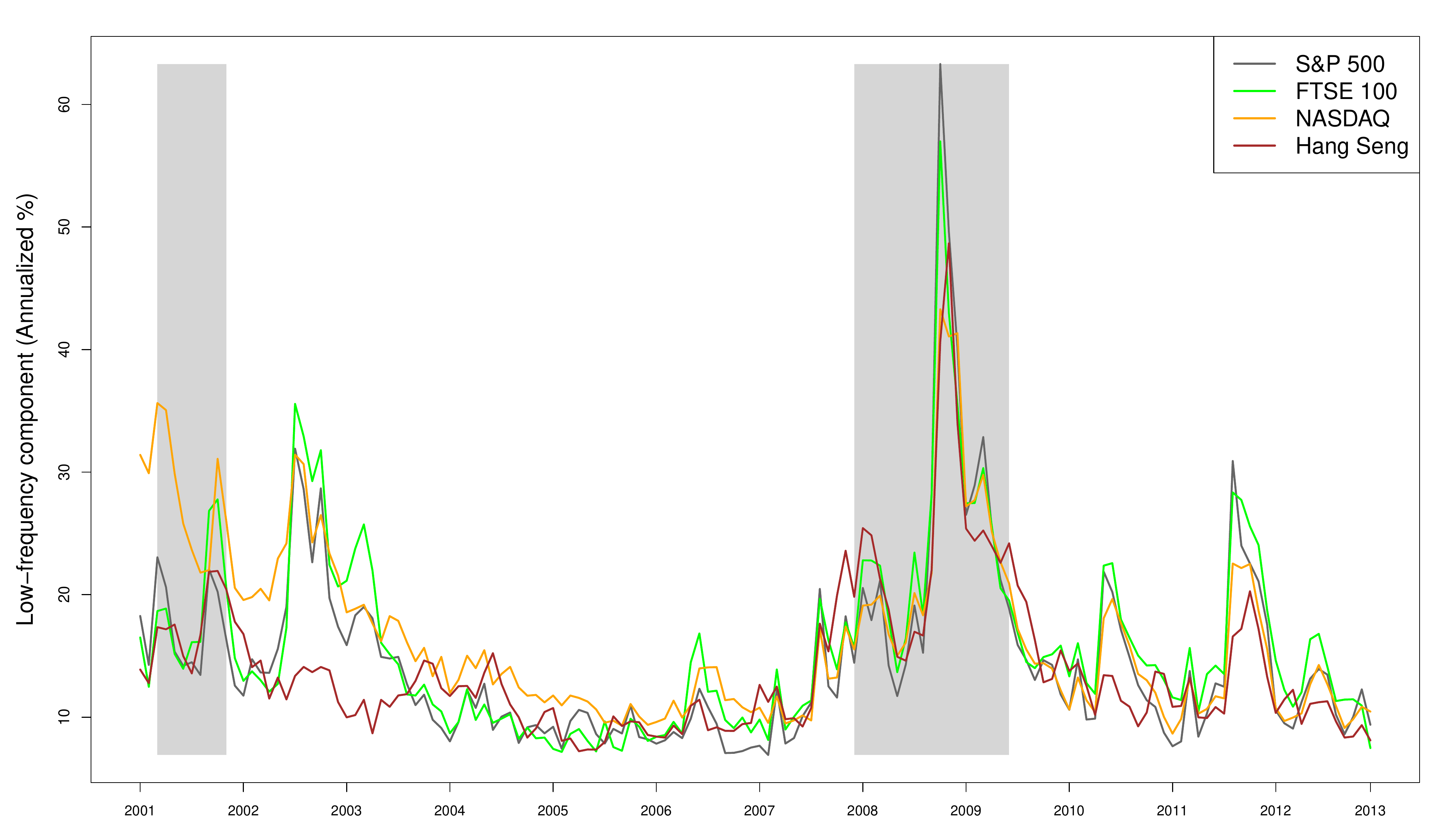}
		\caption{\MEMc\ $\tau_{t}$ \label{fig:cmem_tau_t}}
	\end{subfigure}%
	\hfill
	\begin{subfigure}[b]{1\textwidth}
		\centering
		\includegraphics[width=1\linewidth]{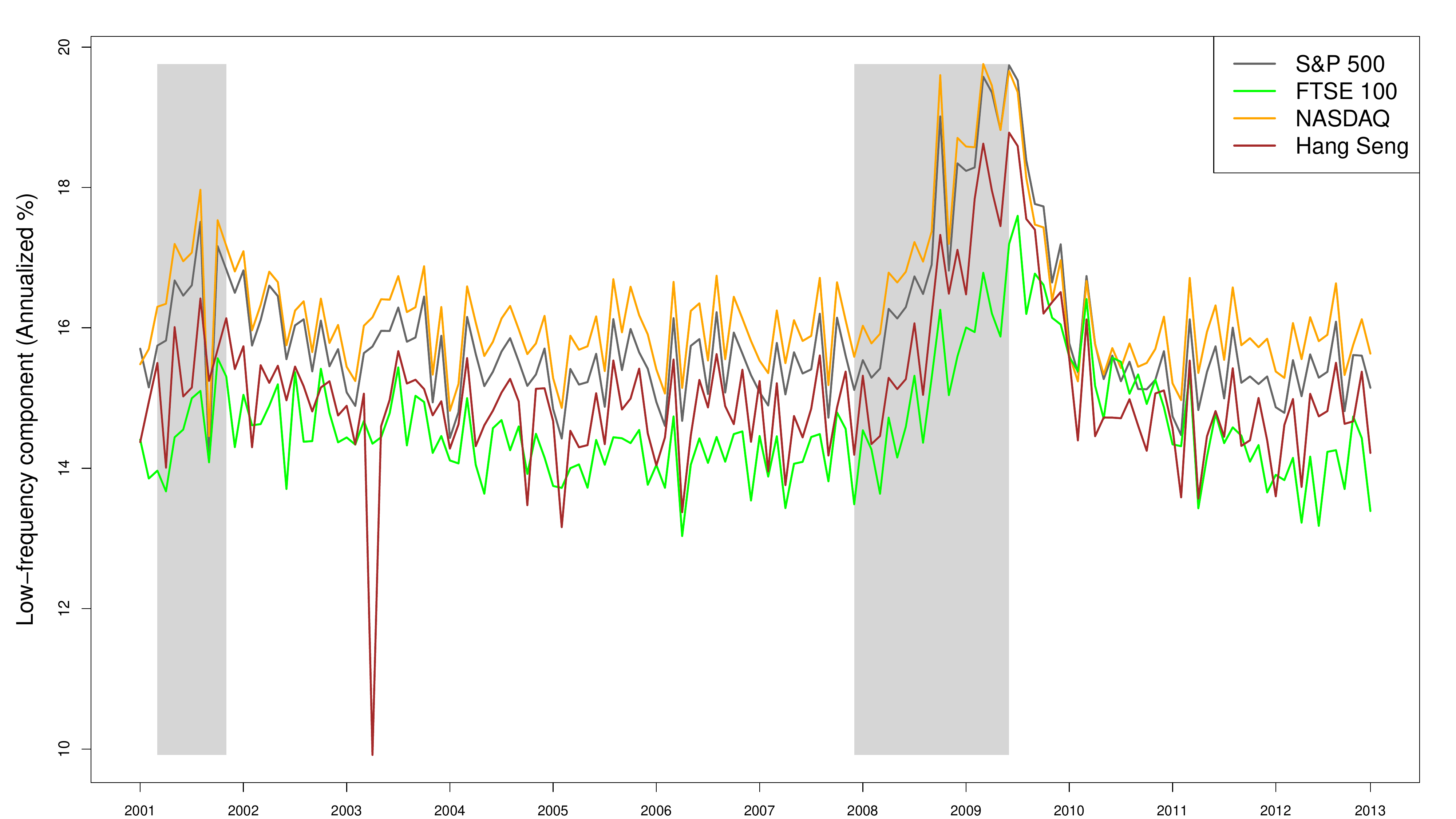}
		\caption{\MEMMIDAS\ $\tau_{t}$ \label{fig:mem_m_tau_t}}
	\end{subfigure}	
	\vspace{-0.1cm}
	\begin{minipage}[t]{1\textwidth}
		\textbf{Notes:} Plot of the \MEMc\  (top plot) and \MEMMIDAS\ (bottom plot) $\tau_{t}$ terms. Shaded areas represent US recession periods (NBER dating).
	\end{minipage}
	
\end{figure}

The two \DMEM\ models produce an estimate of the \lrc\, which is at a daily frequency for the \MEMc, and at a monthly frequency for the \MEMMIDAS: in order for them to be compared, we choose to aggregate the former at the monthly level by averaging to the same scale, with an obvious change of notation for the objects involved, by dropping the subscript $i$. In Figure \ref{fig:cmem_mem_midas_tau} we report the four $\tau_{t}$  components (for each index) estimated with the \MEMc\ (top plot), and  with the \MEMMIDAS\ (bottom plot). It seems that the $\tau_t$ components have a similar pattern across all the indices, within the same specification (more on this later). To investigate this aspect, in Table \ref{tab:cmem_mem_midas_corr}, we report the correlations (numbers in regular text) among the $\tau_t$ terms of the \MEMc\ and among those of the \MEMMIDAS\ (number in \textit{italics}); on the main diagonal, we reproduce the correlation coefficient between $\tau_t$'s estimated by the two different methods: they are all above $0.5$ pointing both to the similarity of the two outcomes, but, by the same token, also to the difference of information and approach used to derive them. As far as the correlations across markets are concerned, neither method delivers consistently higher values than the other. By and large, the commonality in the $\tau_t$'s for different indices is confirmed and, as expected, the values are higher for the two US and the UK markets.\footnote{It is not relevant, for the sake of our argument, to address the issue of the different opening schedules across time zones here.} 

\begin{table}[b]
\centering
\caption{\MEMc\ and \MEMMIDAS. Correlations among the $\tau_t$ components \label{tab:cmem_mem_midas_corr}}
\vspace{-0.35cm}
 \begin{adjustbox}{max height=0.7\textheight , max width=\textwidth}
  \begin{threeparttable}
\begin{tabular}{rrrrr}
\toprule 
 							& S\&P 500  & FTSE 100 	& NASDAQ 	& Hang Seng \\ 
  \midrule
\multirow{2}{*}{S\&P 500} 	& \multirow{2}{*}{\textbf{0.502}}& 0.956 		& 0.854 			& 0.806 		\\ 
		 					& 						& \textit{0.808}& 	\textit{0.983}	& \textit{0.841}		 \\ 
\addlinespace																		
\hdashline																		
\addlinespace 
\multirow{2}{*}{FTSE 100}  	& 			& \multirow{2}{*}{\textbf{0.512}}	& 0.808 	& 0.761 		\\ 
		 					& 			& 			& \textit{0.733}		 	& 	\textit{0.764} 		\\ 
\addlinespace																		
\hdashline																		
\addlinespace 
\multirow{2}{*}{NASDAQ} 	& 			&  			& \multirow{2}{*}{\textbf{0.512}} 	&  0.731	\\ 
							&  			&  			&  			&  	\textit{0.818}		\\ 
							\addlinespace																		
\hdashline																		
\addlinespace 
\multirow{2}{*}{Hang Seng} 	& 			&  			& 			& \multirow{2}{*}{\textbf{0.571}}	 \\ 
							& 			&  			& 			&  		\\

   \bottomrule
\end{tabular}
  \begin{tablenotes}[flushleft]
   \setlength\labelsep{0pt}
   \footnotesize
\item \textbf{Notes}: Numbers in bold are the correlations among the low-frequency terms of the \MEMc\ and \MEMMIDAS\ models. Numbers in regular text and \textit{italics} are the correlations among the indexes for the  \MEMc\  and \MEMMIDAS\ models, respectively.
\end{tablenotes}
\end{threeparttable}
\end{adjustbox}
\end{table}

\subsection{Out-of-sample analysis}

In the out-of-sample exercise, each model is estimated using a rolling window of twelve years (approximately, 3000 daily observations). Subsequently, the one-step-ahead forecasts are generated for the following two months, conditionally on the parameters' estimates previously obtained. Then, the estimation window shifts forward by two months, new out-of-sample forecasts are produced as in the previous step for the following two months, and so forth until the end of the series.\footnote{The results presented here are robust to larger refitting periods. Additional material is available upon request.} 
The first estimation period coincides with the in-sample period 2001--2012. The out-of-sample performances of the models, for each index under consideration, are depicted in Tables \ref{tab:oos_est_sp500_2} to \ref{tab:oos_est_hang_seng_2}. It can be easily noted that the largest gray area (indicating inclusion in the MCS) for all the tables, LFs and out-of-sample periods is for the \MEM--based models, followed by some more scattered presence of the \AHAR. The consistent presence of these models is reassuring in terms of modeling realized volatility directly, on the one hand, and within that class in terms of the convenience to treat innovation terms as entering multiplicatively. Modeling conditional volatility through the conditional second moments of returns seems to be dominated according to either metric in the loss functions. Somewhat disappointingly, \RGARCH\, seldom enters the MCS.

\begin{figure}[hp]
	\centering
	\caption{\MEMc\ and \MEMMIDAS\ out-of-sample volatilities} 
	\label{fig:oos_comp_2}
	\vspace{-0.35cm}
	\begin{subfigure}[b]{0.49\textwidth}
		\centering
		\includegraphics[width=1\linewidth]{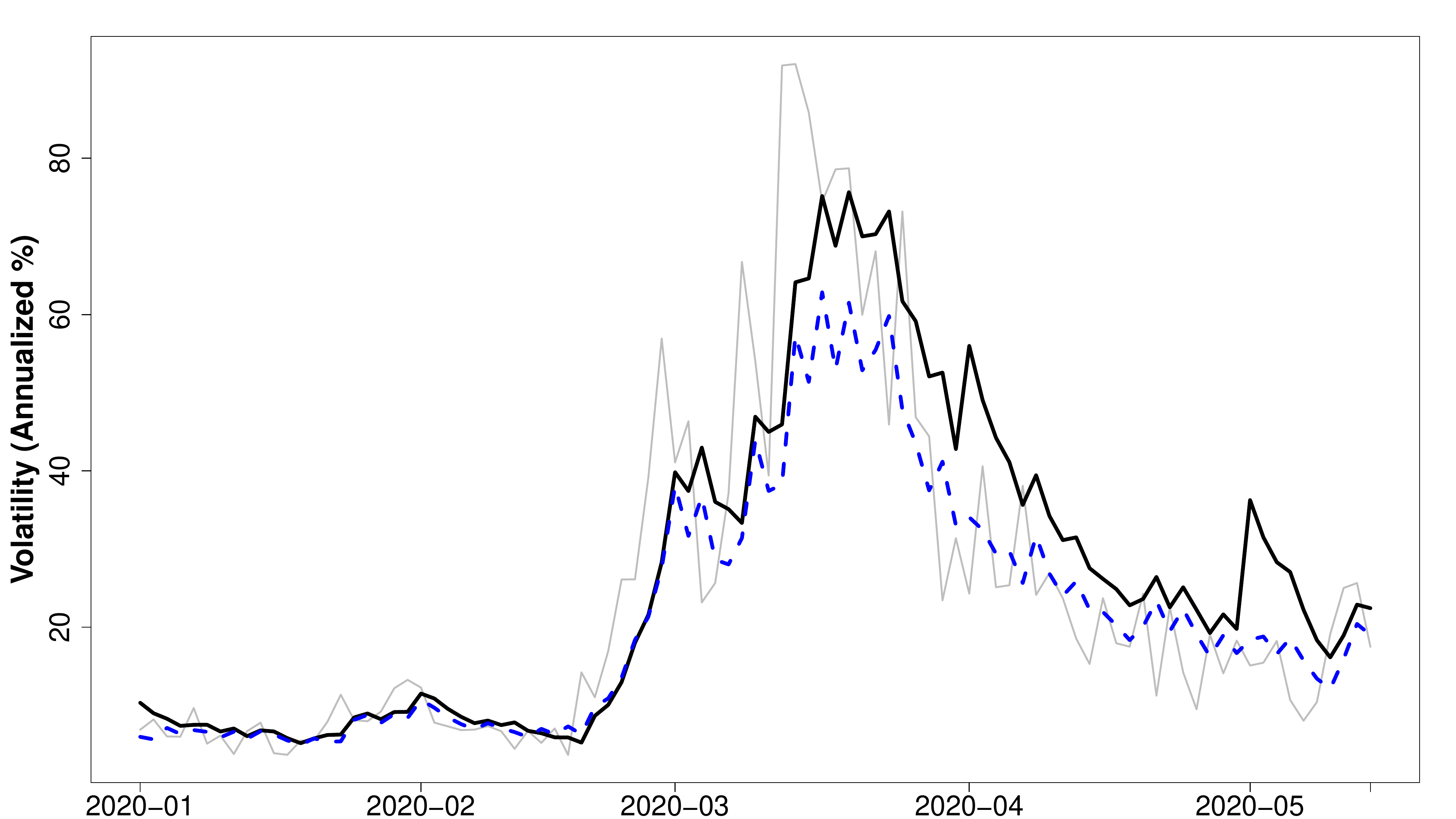}
		\caption{S\&P 500 \label{fig:sp500_oos_2_months}}
	\end{subfigure}%
	%\hfill
	\begin{subfigure}[b]{0.49\textwidth}
		\centering
		\includegraphics[width=1\linewidth]{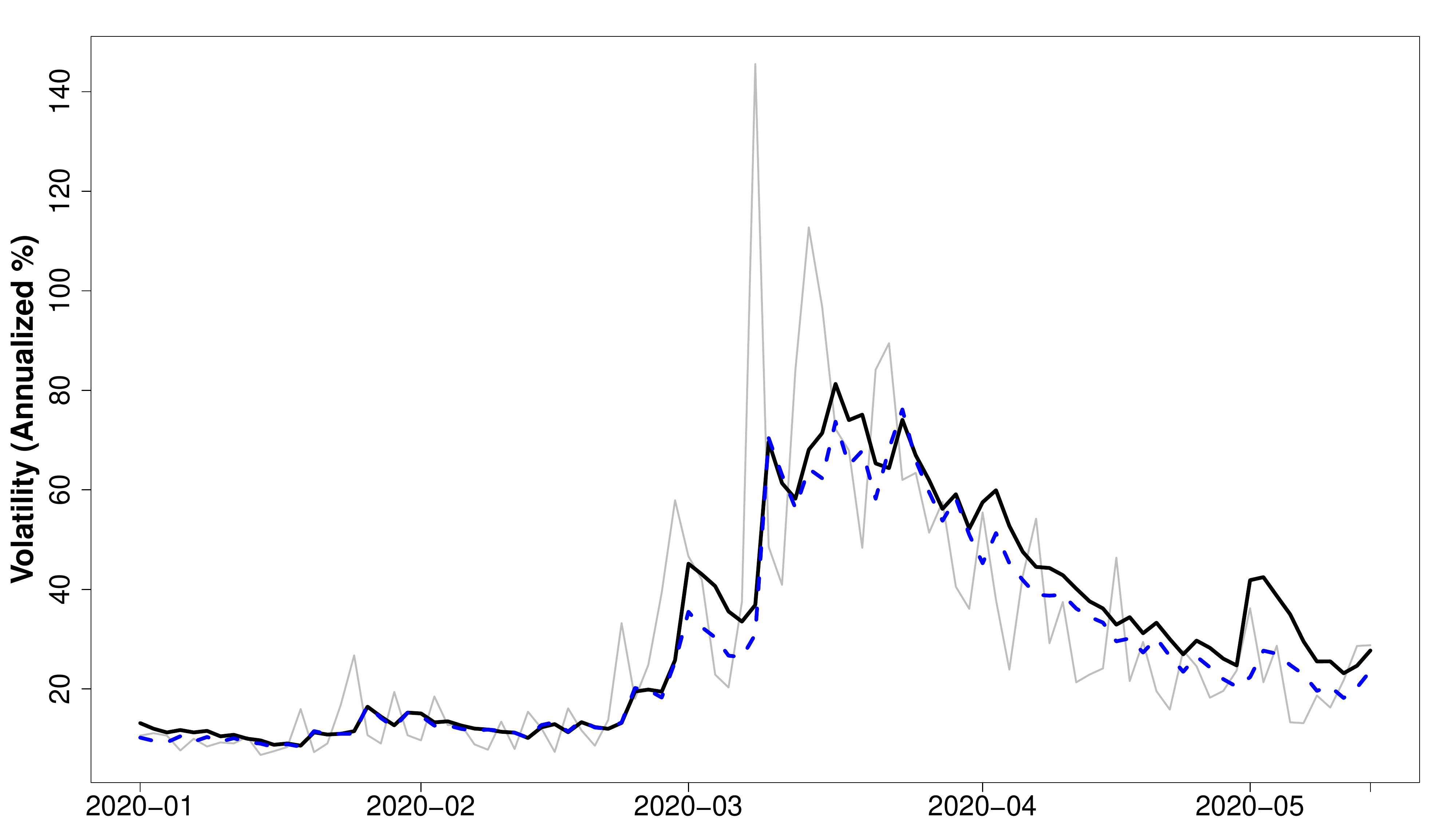}
		\caption{FTSE 100 \label{fig:ftse_100_oos_2_months}}
	\end{subfigure}
	\vskip\baselineskip
	\begin{subfigure}[b]{0.49\textwidth}
		\centering
		\includegraphics[width=1\linewidth]{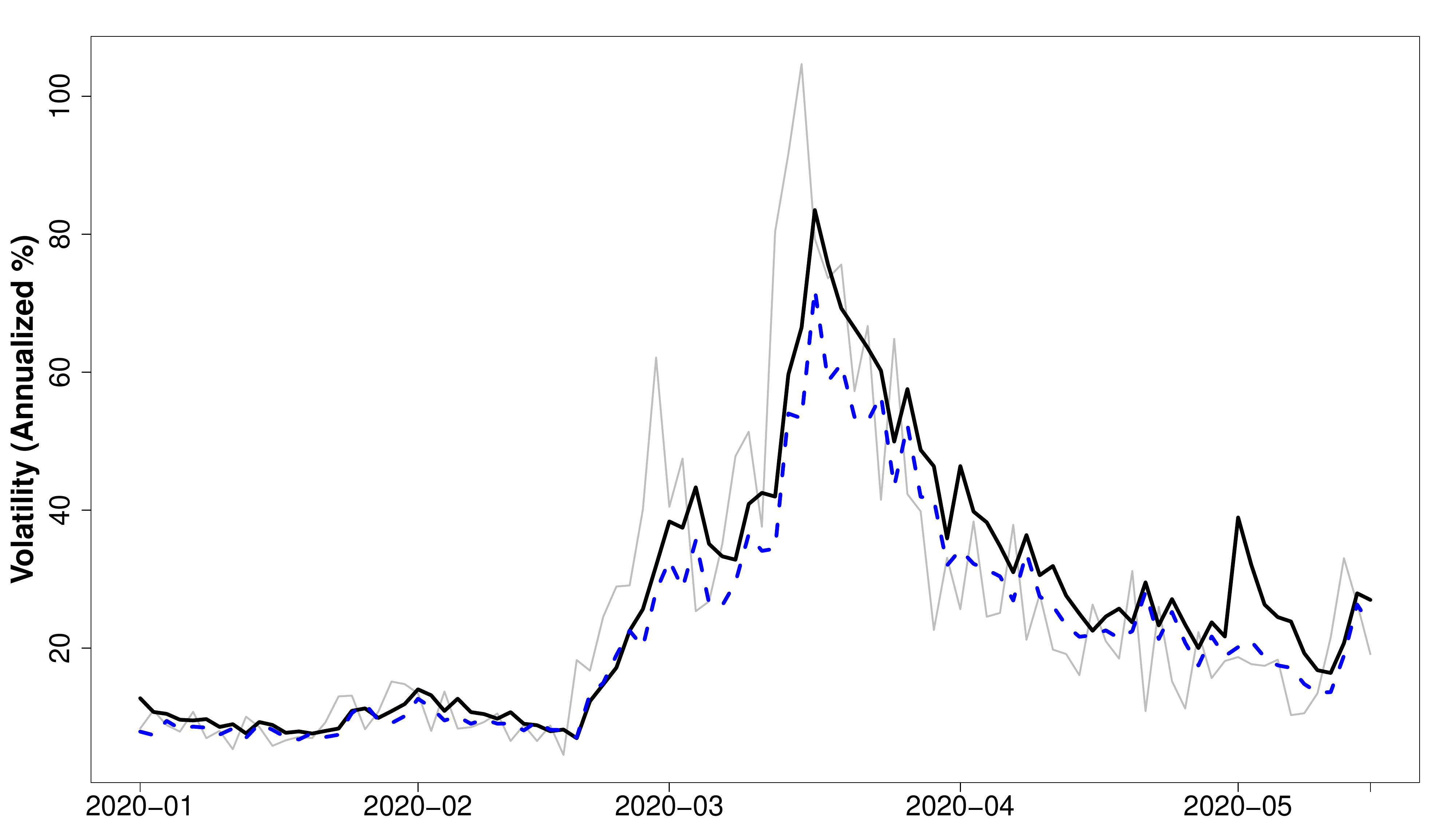}
		\caption{NASDAQ \label{fig:nasdaq_oos_2_months}}
	\end{subfigure}
	%\hfill
	\begin{subfigure}[b]{0.49\textwidth}
		\centering
		\includegraphics[width=1\linewidth]{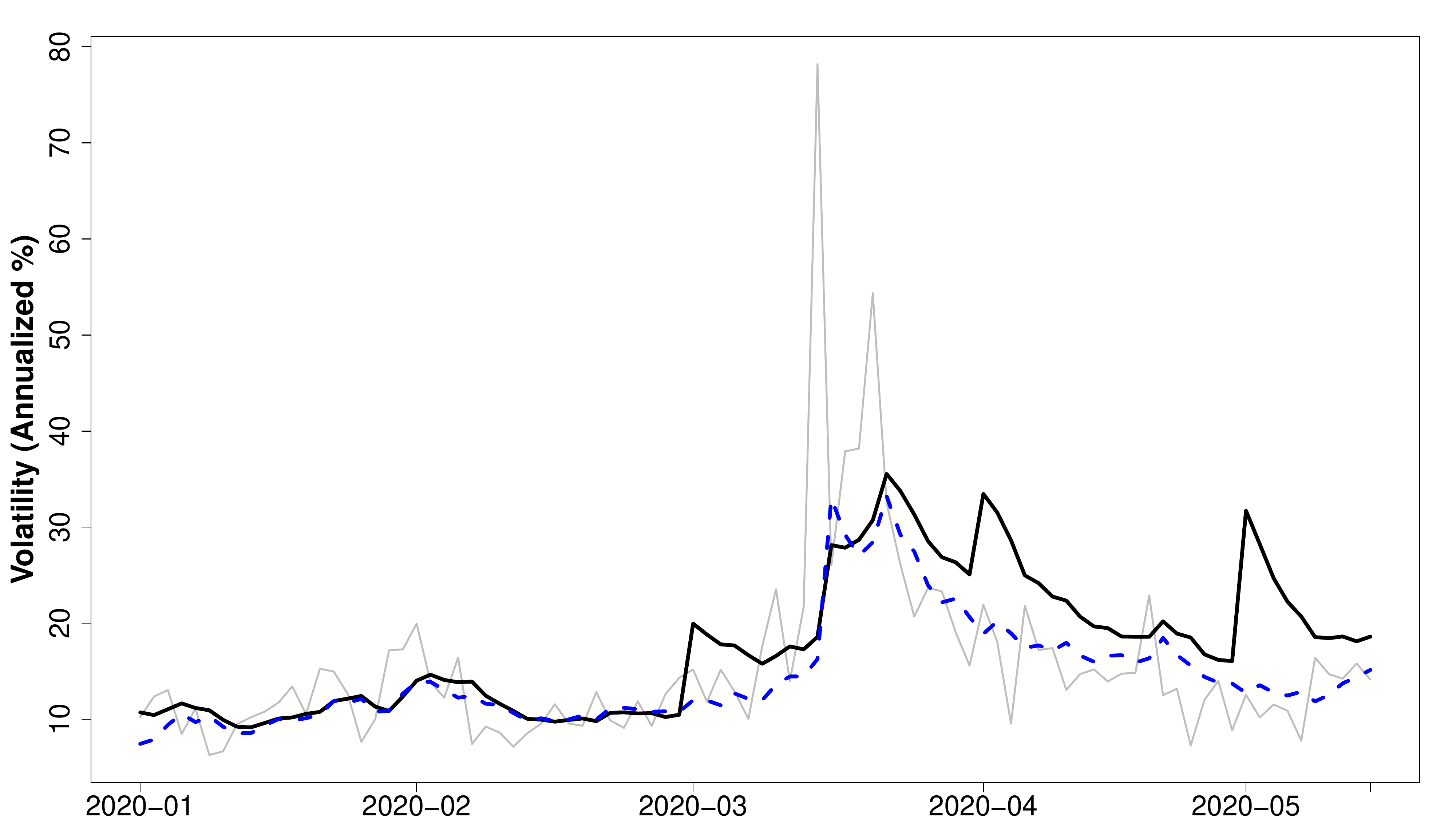}
		\caption{Hang Seng \label{fig:hang_seng_oos_2_months}}
	\end{subfigure}%
		\vspace{-0.1cm}
	\begin{minipage}[t]{1\textwidth}
		\textbf{Notes:} Realized kernel volatility (grey line),  \MEMc\ (blue dotted line) and \MEMMIDAS\ (black line) out-of-sample estimated volatilities. Period: 2 January 2020 - 15 May 2020.
	\end{minipage}
	
\end{figure}

To gain some further insights as of the behavior of each model in relationship with the observed volatility pattern, we suggest a graphical comparison (Figure \ref{fig:oos_comp_2}) between the two \DMEM\, models introduced in this paper. To that end, we reproduce, for the last period of our sample (from 2 January 2020 to 15 May 2020), the out--of--sample forecasts next to the realized kernel volatility.

\begin{table}[htbp]
\centering
\caption{Out-of-sample comparison. S\&P 500 \label{tab:oos_est_sp500_2}}
\vspace{-0.35cm}
 \begin{adjustbox}{max height=0.7\textheight , max width=\textwidth}
  \begin{threeparttable}
\begin{tabular}{l ........}
\toprule  
\textbf{}               	&	  \mc{\AMEM}         &	       \mc{\MEMc}   	&	   \mc{\MEMMIDAS}          	&	       \mc{\AHAR}            	     	&	  \mc{\GJR}   &       \mc{\GM}   & 	         \mc{\DAGM}       &	         \mc{\RGARCH}       \\
\midrule                                                                                                          
  \multicolumn{9}{c}{QLIKE}\\
  \addlinespace
  2013 & \cellcolor{gray!75}0.081 & \cellcolor{gray!75}0.08 & 0.084 & 0.087 & 0.115 & 0.131 & 0.15 & 0.105 \\ 
   2014 & 0.074 & \cellcolor{gray!75}0.068 & 0.074 & 0.082 & 0.124 & 0.143 & 0.181 & 0.097 \\ 
   2015 & \cellcolor{gray!75}0.061 & \cellcolor{gray!75}0.061 & \cellcolor{gray!75}0.061 & \cellcolor{gray!75}0.064 & 0.086 & 0.123 & 0.088 & 0.086 \\ 
   2016 & 0.064 & 0.068 & \cellcolor{gray!75}0.062 & 0.068 & 0.096 & 0.119 & 0.12 & 0.088 \\ 
   2017 & \cellcolor{gray!75}0.072 & \cellcolor{gray!75}0.072 & \cellcolor{gray!75}0.074 & 0.09 & 0.169 & 0.227 & 0.176 & 0.106 \\ 
   2018 & \cellcolor{gray!75}0.06 & \cellcolor{gray!75}0.055 & \cellcolor{gray!75}0.061 & \cellcolor{gray!75}0.062 & 0.093 & 0.091 & 0.102 & 0.075 \\ 
   2019 & \cellcolor{gray!75}0.07 & \cellcolor{gray!75}0.066 & 0.073 & 0.076 & 0.098 & 0.127 & 0.116 & 0.092 \\ 
   2020 & \cellcolor{gray!75}0.072 & \cellcolor{gray!75}0.07 & \cellcolor{gray!75}0.086 & \cellcolor{gray!75}0.11 & \cellcolor{gray!75}0.083 & 0.145 & 0.175 & \cellcolor{gray!75}0.077 \\ 
   Full & 0.069 & \cellcolor{gray!75}0.067 & 0.071 & 0.077 & 0.11 & 0.138 & 0.135 & 0.092 \\ 
   
     \addlinespace
 \hdashline
   \addlinespace
     \multicolumn{9}{c}{MSE}\\
  \addlinespace
  2013 & \cellcolor{gray!75}0.051 & \cellcolor{gray!75}0.051 & \cellcolor{gray!75}0.052 & 0.059 & 0.082 & 0.098 & 0.124 & 0.075 \\ 
   2014 & 0.044 & \cellcolor{gray!75}0.041 & \cellcolor{gray!75}0.043 & 0.049 & 0.087 & 0.102 & 0.124 & 0.064 \\ 
   2015 & \cellcolor{gray!75}0.098 & \cellcolor{gray!75}0.094 & \cellcolor{gray!75}0.097 & \cellcolor{gray!75}0.1 & 0.128 & 0.137 & \cellcolor{gray!75}0.12 & 0.136 \\ 
   2016 & \cellcolor{gray!75}0.058 & \cellcolor{gray!75}0.066 & \cellcolor{gray!75}0.059 & \cellcolor{gray!75}0.064 & 0.088 & 0.108 & 0.134 & 0.087 \\ 
   2017 & \cellcolor{gray!75}0.016 & \cellcolor{gray!75}0.017 & \cellcolor{gray!75}0.016 & 0.025 & 0.056 & 0.097 & 0.068 & 0.029 \\ 
   2018 & 0.091 & \cellcolor{gray!75}0.082 & 0.089 & \cellcolor{gray!75}0.087 & 0.147 & 0.125 & 0.135 & 0.118 \\ 
   2019 & \cellcolor{gray!75}0.046 & \cellcolor{gray!75}0.042 & \cellcolor{gray!75}0.047 & 0.051 & 0.076 & 0.1 & 0.088 & 0.071 \\ 
  2020 & \cellcolor{gray!75}0.462 & \cellcolor{gray!75}0.533 & \cellcolor{gray!75}0.575 & \cellcolor{gray!75}0.58 & \cellcolor{gray!75}0.488 & 0.772 & 1.058 & \cellcolor{gray!75}0.52 \\ 
  Full & \cellcolor{gray!75}0.078 & \cellcolor{gray!75}0.08 & \cellcolor{gray!75}0.084 & 0.088 & 0.114 & 0.143 & 0.161 & 0.105 \\ 
  
\bottomrule
\end{tabular}
  \begin{tablenotes}[flushleft]
   \setlength\labelsep{0pt}
   \footnotesize
\item \textbf{Notes}:  The table reports the averages
 of the QLIKE and MSE loss functions. Rolling window: twelve years. Refitting frequency: two months. Shades of gray denote inclusion in the MCS at significance level $\alpha=0.25$.
\end{tablenotes}
\end{threeparttable}
\end{adjustbox}
\end{table}
\begin{table}[htbp]
\centering
\caption{Out-of-sample comparison. FTSE 100 \label{tab:oos_est_ftse_100_2}}
\vspace{-0.35cm}
 \begin{adjustbox}{max height=0.7\textheight , max width=\textwidth}
   \begin{threeparttable}
\begin{tabular}{l ........}
\toprule  
\textbf{}               	&	  \mc{\AMEM}         &	       \mc{\MEMc}   	&	   \mc{\MEMMIDAS}          	&	       \mc{\AHAR}            	     	&	  \mc{\GJR}   &       \mc{\GM}   & 	         \mc{\DAGM}       &	         \mc{\RGARCH}       \\
\midrule                                                                                                          
  \multicolumn{9}{c}{QLIKE}\\
  \addlinespace
 2013 & 0.053 & \cellcolor{gray!75}0.05 & 0.058 & 0.056 & 0.058 & 0.066 & 0.073 & 0.068 \\ 
   2014 & \cellcolor{gray!75}0.046 & \cellcolor{gray!75}0.046 & \cellcolor{gray!75}0.047 & 0.055 & 0.059 & 0.091 & 0.135 & 0.054 \\ 
   2015 & \cellcolor{gray!75}0.053 & \cellcolor{gray!75}0.054 & \cellcolor{gray!75}0.054 & 0.06 & 0.062 & 0.061 & \cellcolor{gray!75}0.061 & 0.061 \\ 
 2016 & \cellcolor{gray!75}0.068 & 0.079 & \cellcolor{gray!75}0.067 & \cellcolor{gray!75}0.072 & \cellcolor{gray!75}0.07 & \cellcolor{gray!75}0.072 & \cellcolor{gray!75}0.066 & \cellcolor{gray!75}0.073 \\ 
  2017 & \cellcolor{gray!75}0.045 & \cellcolor{gray!75}0.045 & \cellcolor{gray!75}0.045 & 0.056 & 0.071 & 0.099 & 0.078 & 0.057 \\ 
  2018 & 0.061 & \cellcolor{gray!75}0.057 & 0.062 & 0.065 & 0.086 & 0.089 & 0.097 & 0.065 \\ 
2019 & \cellcolor{gray!75}0.037 & \cellcolor{gray!75}0.037 & \cellcolor{gray!75}0.039 & 0.044 & 0.054 & 0.07 & 0.065 & 0.049 \\ 
 2020 & \cellcolor{gray!75}0.085 & \cellcolor{gray!75}0.092 & \cellcolor{gray!75}0.092 & \cellcolor{gray!75}0.132 & \cellcolor{gray!75}0.098 & 0.15 & 0.515 & \cellcolor{gray!75}0.094 \\ 
Full & \cellcolor{gray!75}0.054 & \cellcolor{gray!75}0.055 & 0.055 & 0.062 & 0.067 & 0.082 & 0.104 & 0.063 \\ 
  
     \addlinespace
 \hdashline
   \addlinespace
     \multicolumn{9}{c}{MSE}\\
  \addlinespace
 2013 & 0.056 & \cellcolor{gray!75}0.053 & 0.058 & 0.061 & 0.068 & 0.076 & 0.084 & 0.081 \\ 
 2014 & \cellcolor{gray!75}0.046 & \cellcolor{gray!75}0.045 & \cellcolor{gray!75}0.046 & 0.052 & 0.057 & 0.108 & 0.09 & 0.054 \\ 
2015 & \cellcolor{gray!75}0.11 & \cellcolor{gray!75}0.111 & \cellcolor{gray!75}0.111 & \cellcolor{gray!75}0.119 & 0.139 & \cellcolor{gray!75}0.12 & \cellcolor{gray!75}0.12 & 0.13 \\ 
  2016 & \cellcolor{gray!75}0.338 & 0.394 & \cellcolor{gray!75}0.336 & \cellcolor{gray!75}0.355 & \cellcolor{gray!75}0.333 & \cellcolor{gray!75}0.332 & \cellcolor{gray!75}0.321 & \cellcolor{gray!75}0.336 \\ 
  2017 & \cellcolor{gray!75}0.028 & \cellcolor{gray!75}0.029 & \cellcolor{gray!75}0.028 & 0.037 & 0.053 & 0.084 & 0.06 & 0.038 \\ 
   2018 & 0.085 & \cellcolor{gray!75}0.081 & 0.085 & 0.088 & 0.129 & 0.127 & 0.14 & 0.091 \\ 
  2019 & \cellcolor{gray!75}0.038 & \cellcolor{gray!75}0.037 & \cellcolor{gray!75}0.039 & 0.045 & 0.07 & 0.092 & 0.085 & 0.052 \\ 
2020 & \cellcolor{gray!75}1.061 & \cellcolor{gray!75}1.135 & \cellcolor{gray!75}1.136 & \cellcolor{gray!75}1.259 & \cellcolor{gray!75}1.14 & 1.617 & 2.341 & \cellcolor{gray!75}1.186 \\ 
  Full & \cellcolor{gray!75}0.148 & 0.159 & 0.152 & 0.166 & 0.172 & 0.208 & 0.239 & 0.165 \\

\bottomrule
\end{tabular}
  \begin{tablenotes}[flushleft]
   \setlength\labelsep{0pt}
   \footnotesize
\item \textbf{Notes}:  The table reports the averages
 of the QLIKE and MSE loss functions. Rolling window: twelve years. Refitting frequency: two months. Shades of gray denote inclusion in the MCS at significance level $\alpha=0.25$.
\end{tablenotes}
\end{threeparttable}
\end{adjustbox}
\end{table}
\begin{table}[htbp]
\centering
\caption{Out-of-sample comparison. NASDAQ \label{tab:oos_est_nasdaq_2}}
\vspace{-0.35cm}
 \begin{adjustbox}{max height=0.7\textheight , max width=\textwidth}
  \begin{threeparttable}
\begin{tabular}{l ........}
\toprule  
\textbf{}               	&	  \mc{\AMEM}         &	       \mc{\MEMc}   	&	   \mc{\MEMMIDAS}          	&	       \mc{\AHAR}            	     	&	  \mc{\GJR}   &       \mc{\GM}   & 	         \mc{\DAGM}       &	         \mc{\RGARCH}       \\
\midrule                                                                                                          
  \multicolumn{9}{c}{Asymmetric LF, under prediction version: QLIKE ($b=-2$)}\\
  \addlinespace
 2013 & \cellcolor{gray!75}0.049 & \cellcolor{gray!75}0.046 & 0.059 & 0.059 & 0.086 & 0.189 & 0.22 & 0.065 \\ 
  2014 & \cellcolor{gray!75}0.058 & \cellcolor{gray!75}0.055 & \cellcolor{gray!75}0.057 & \cellcolor{gray!75}0.059 & 0.09 & 0.107 & 0.12 & 0.098 \\ 
 2015 & \cellcolor{gray!75}0.054 & \cellcolor{gray!75}0.051 & \cellcolor{gray!75}0.054 & \cellcolor{gray!75}0.053 & 0.079 & 0.103 & 0.08 & 0.071 \\ 
  2016 & \cellcolor{gray!75}0.052 & \cellcolor{gray!75}0.052 & \cellcolor{gray!75}0.051 & \cellcolor{gray!75}0.054 & 0.07 & 0.122 & 0.068 & 0.065 \\ 
   2017 & \cellcolor{gray!75}0.071 & \cellcolor{gray!75}0.071 & \cellcolor{gray!75}0.071 & \cellcolor{gray!75}0.077 & 0.131 & 0.152 & 0.133 & 0.09 \\ 
 2018 & \cellcolor{gray!75}0.054 & \cellcolor{gray!75}0.052 & \cellcolor{gray!75}0.055 & \cellcolor{gray!75}0.061 & 0.085 & 0.072 & 0.108 & 0.061 \\ 
  2019 & \cellcolor{gray!75}0.066 & \cellcolor{gray!75}0.063 & \cellcolor{gray!75}0.067 & 0.073 & 0.09 & 0.096 & 0.093 & 0.076 \\ 
  2020 & \cellcolor{gray!75}0.062 & \cellcolor{gray!75}0.069 & \cellcolor{gray!75}0.071 & \cellcolor{gray!75}0.113 & \cellcolor{gray!75}0.096 & 0.162 & 0.246 & \cellcolor{gray!75}0.063 \\ 
Full & 0.058 & \cellcolor{gray!75}0.056 & 0.06 & 0.065 & 0.09 & 0.122 & 0.124 & 0.075 \\

     \addlinespace
 \hdashline
   \addlinespace
     \multicolumn{9}{c}{Symmetric LF: MSE ($b=0$)}\\
  \addlinespace
  2013 & \cellcolor{gray!75}0.03 & \cellcolor{gray!75}0.029 & 0.047 & 0.041 & 0.065 & 0.398 & 0.483 & 0.045 \\ 
 2014 & \cellcolor{gray!75}0.052 & \cellcolor{gray!75}0.049 & \cellcolor{gray!75}0.051 & \cellcolor{gray!75}0.053 & 0.095 & 0.115 & 0.137 & 0.113 \\ 
  2015 & 0.107 & \cellcolor{gray!75}0.101 & 0.106 & \cellcolor{gray!75}0.104 & 0.14 & 0.168 & 0.137 & 0.132 \\ 
  2016 & \cellcolor{gray!75}0.056 & \cellcolor{gray!75}0.058 & \cellcolor{gray!75}0.055 & \cellcolor{gray!75}0.057 & 0.07 & 0.091 & 0.068 & 0.074 \\ 
   2017 & \cellcolor{gray!75}0.03 & \cellcolor{gray!75}0.032 & \cellcolor{gray!75}0.03 & 0.036 & 0.076 & 0.092 & 0.07 & 0.044 \\ 
  2018 & \cellcolor{gray!75}0.108 & \cellcolor{gray!75}0.105 & \cellcolor{gray!75}0.108 & \cellcolor{gray!75}0.109 & 0.176 & 0.139 & 0.161 & 0.125 \\ 
   2019 & \cellcolor{gray!75}0.06 & \cellcolor{gray!75}0.058 & \cellcolor{gray!75}0.061 & 0.067 & 0.093 & 0.092 & 0.089 & 0.075 \\ 
   2020 & \cellcolor{gray!75}0.426 & 0.545 & 0.49 & 0.577 & 0.823 & 1.202 & 3.296 & \cellcolor{gray!75}0.443 \\ 
Full& \cellcolor{gray!75}0.082 & \cellcolor{gray!75}0.086 & \cellcolor{gray!75}0.087 & 0.093 & 0.139 & 0.21 & 0.323 & 0.105 \\ 

\bottomrule
\end{tabular}
  \begin{tablenotes}[flushleft]
   \setlength\labelsep{0pt}
   \footnotesize
\item \textbf{Notes}:  The table reports the averages
 of the QLIKE and MSE loss functions. Rolling window: twelve years. Refitting frequency: two months. Shades of gray denote inclusion in the MCS at significance level $\alpha=0.25$.
\end{tablenotes}
\end{threeparttable}
\end{adjustbox}
\end{table}
\begin{table}[htbp]
\centering
\caption{Out-of-sample comparison. Hang Seng \label{tab:oos_est_hang_seng_2}}
\vspace{-0.35cm}
 \begin{adjustbox}{max height=0.7\textheight , max width=\textwidth}
  \begin{threeparttable}
\begin{tabular}{l ........}
\toprule  
\textbf{}               	&	  \mc{\AMEM}         &	       \mc{\MEMc}   	&	   \mc{\MEMMIDAS}          	&	       \mc{\AHAR}            	     	&	  \mc{\GJR}   &       \mc{\GM}   & 	         \mc{\DAGM}       &	         \mc{\RGARCH}       \\
\midrule                                                                                                          
  \multicolumn{9}{c}{QLIKE}\\
  \addlinespace
 2013 & \cellcolor{gray!75}0.058 & \cellcolor{gray!75}0.059 & \cellcolor{gray!75}0.059 & 0.063 & 0.072 & 0.086 & 0.114 & 0.066 \\ 
   2014 & \cellcolor{gray!75}0.049 & \cellcolor{gray!75}0.051 & \cellcolor{gray!75}0.051 & 0.055 & 0.061 & 0.065 & 0.086 & 0.057 \\ 
   2015 & \cellcolor{gray!75}0.078 & \cellcolor{gray!75}0.075 & \cellcolor{gray!75}0.077 & \cellcolor{gray!75}0.076 & 0.086 & 0.09 & 0.09 & 0.084 \\ 
   2016 & 0.05 & 0.054 & \cellcolor{gray!75}0.049 & 0.056 & 0.061 & 0.07 & 0.064 & 0.06 \\ 
   2017 & \cellcolor{gray!75}0.049 & \cellcolor{gray!75}0.049 & \cellcolor{gray!75}0.051 & 0.056 & 0.065 & 0.161 & 0.117 & 0.062 \\ 
   2018 & 0.042 & \cellcolor{gray!75}0.04 & 0.045 & 0.046 & 0.051 & 0.044 & 0.078 & 0.047 \\ 
   2019 & \cellcolor{gray!75}0.048 & \cellcolor{gray!75}0.047 & \cellcolor{gray!75}0.049 & 0.05 & 0.06 & 0.073 & 0.07 & 0.058 \\ 
  2020 & \cellcolor{gray!75}0.061 & \cellcolor{gray!75}0.063 & 0.083 & \cellcolor{gray!75}0.061 & 0.098 & 0.176 & 0.126 & \cellcolor{gray!75}0.067 \\ 
 Full & \cellcolor{gray!75}0.054 & \cellcolor{gray!75}0.054 & 0.056 & 0.058 & 0.067 & 0.089 & 0.091 & 0.062 \\ 
    
 \addlinespace
 \hdashline
   \addlinespace
     \multicolumn{9}{c}{MSE}\\
  \addlinespace
  2013 & \cellcolor{gray!75}0.052 & \cellcolor{gray!75}0.052 & \cellcolor{gray!75}0.053 & 0.057 & 0.065 & 0.08 & 0.121 & 0.06 \\ 
   2014 & \cellcolor{gray!75}0.04 & \cellcolor{gray!75}0.041 & \cellcolor{gray!75}0.041 & 0.046 & 0.053 & 0.058 & 0.085 & 0.049 \\ 
   2015 & \cellcolor{gray!75}0.184 & \cellcolor{gray!75}0.174 & \cellcolor{gray!75}0.181 & \cellcolor{gray!75}0.175 & 0.203 & 0.205 & 0.207 & 0.197 \\ 
   2016 & 0.077 & 0.083 & \cellcolor{gray!75}0.076 & 0.085 & 0.093 & 0.104 & 0.096 & 0.091 \\ 
   2017 & \cellcolor{gray!75}0.028 & \cellcolor{gray!75}0.028 & \cellcolor{gray!75}0.029 & 0.034 & 0.042 & 0.059 & 0.06 & 0.038 \\ 
   2018 & 0.06 & \cellcolor{gray!75}0.057 & 0.063 & 0.064 & 0.074 & 0.062 & 0.072 & 0.07 \\ 
   2019 & \cellcolor{gray!75}0.046 & \cellcolor{gray!75}0.046 & \cellcolor{gray!75}0.047 & \cellcolor{gray!75}0.049 & 0.06 & 0.076 & 0.074 & 0.058 \\ 
   2020 & \cellcolor{gray!75}0.261 & \cellcolor{gray!75}0.257 & 0.342 & \cellcolor{gray!75}0.254 & 0.346 & 0.815 & 0.46 & 0.28 \\ 
 Full & \cellcolor{gray!75}0.079 & \cellcolor{gray!75}0.078 & 0.084 & 0.082 & 0.097 & 0.128 & 0.12 & 0.091 \\ 
  
\bottomrule
\end{tabular}
  \begin{tablenotes}[flushleft]
   \setlength\labelsep{0pt}
   \footnotesize
\item \textbf{Notes}: The table reports the averages
 of the QLIKE and MSE loss functions. Rolling window: twelve years. Refitting frequency: two months. Shades of gray denote inclusion in the MCS at significance level $\alpha=0.25$.
\end{tablenotes}
\end{threeparttable}
\end{adjustbox}
\end{table}

%\input{figure_oos_vol_6}

%\section*{References}

\section{Concluding Remarks}
\label{sect:Concl}

Two different general approaches can be followed when forecasting asset return volatility: one is the GARCH approach where the conditional variance is estimated from return data, the other is modeling the conditional expectation of volatility using ultra--high frequency measures of realized volatility data. In the first approach, therefore, measurement and modeling are comprised within the same framework, while, in the second, the two aspects are decoupled. The merits of the GARCH model are testified by the hundreds of thousands of theoretical and empirical contributions since the seminal paper by \cite{Engle:1982}. This type of approach has been enriched over the years by successive refinements, with the goal to capture some empirical regularities in the pattern of the observed time series. This is the case for the consideration of a time--varying local average  in the conditional variance, a feature addressed by \cite{Engle:Rangel:2008}, also in reference to its economic interpretation to macro economic fluctuations. As a parallel approach, direct modeling of realized measures of volatility has the advantage to exploit the better theoretical properties of these ultra--high frequency measures (less noisy than squared returns).  

For either approach, the consideration of how complicated it is to collect the data and to fine-tune a model to derive the forecast has to be weighed against the actual reward in an improved forecasting performance. The availability of freely downloadable price data still maintains popularity with the \GARCH\, approach (especially among practitioners), but it is also true that the number of high--frequency data vendors is expanding and that DIY processing and storing tick-by-tick data is not a prohibitive task. 

A comparison across models can be interpreted as an exercise that aims at assessing the capability of each model to reproduce empirical regularities in the data, but also at establishing how important those stylized facts are when taken on an out--of--sample terrain. 

In this context, our paper has two clear outcomes: one is to suggest that modeling realized volatility delivers better results than going through a \GARCH--type approach; the second is to show that incorporating the feature that average volatility by subperiod is time--varying provides an advantage in forecasting. For the first outcome, there are clear merits in using a model in which the errors enter multiplicatively, as in the \MEM: this mitigates the attenuation bias in realized volatility models as documented by \cite{Cipollini:Gallo:Otranto:2020}, because it takes into explicit consideration the heteroskedastic nature of volatility measurement errors. For the second outcome, we suggest that doubling the multiplicative components incorporating a slow moving and a \src\ components of volatility dynamics delivers better results, at least for our four stock market indices. We contributed two such models, differentiated by the type of information entering the low--frequency component: in the \MEMc, we use the same daily data, but we allow for a more persistent dynamics; in the \MEMMIDAS, we use a monthly macro-variable (the US industrial production) the variations of which combine in a smooth component which exploits the mixed sampling results by \cite{Ghysels:SantaClara:Valkanov:2006} and by \cite{Engle:Ghysels:Sohn:2013}. While our \MEMMIDAS\ performs better  than the corresponding \GM\, or \DAGM\, in a \GARCH\, context, its delivering a $\tau_t$ which lags behind relative to the bursts of volatility makes it, at times, preferred by another member of the \DMEM\, family, namely the \MEMc. We can see a convenience in using the \MEMMIDAS within a scenario--type approach designing prolonged periods of downturns in economic activity (not necessarily limited to our choice of US industrial production): the impact and aftermath of the COVID--19 health emergency on the financial volatility may thus be studied in projecting to the medium term this channel of transmission originating in the real economy.

While refinements are still possible (e.g. the use of a second lag in making use of observed volatility values, or a \DAGM\, extension within the \MEMMIDAS), one indication that emerges from the empirical results is that the components estimated by our models have some commonality that should be exploited -- in a common factor sense -- by a joint modeling of the series.

\clearpage
% \bibliographystyle{chicago}
% \bibliographystyle{apecon}
% \bibliography{BIBLIO}

\begin{thebibliography}{38}
\providecommand{\natexlab}[1]{#1}

\bibitem[{Amado \textit{et~al.}(2019)Amado, Silvennoinen and
  Ter{\"{a}}svirta}]{Amado:Silva:Terasvirta:2019}
Amado, C., Silvennoinen, A. and Ter{\"{a}}svirta, T. (2019) Models with
  multiplicative decomposition of conditional variances and correlations, in
  \textit{Financial Mathematics, Volatility and Covariance Modelling} (Eds.)
  J.~Chevallier, S.~Goutte, D.~Guerreiro, S.~Saglio and B.~Sanhaji, Routledge,
  vol.~2.

\bibitem[{Amado and Ter{\"{a}}svirta(2008)}]{Amado:Terasvirta:2008}
Amado, C. and Ter{\"{a}}svirta, T. (2008) Modelling conditional and
  unconditional heteroskedasticity with smoothly time-varying structure, Tech.
  Rep.~8, CREATES Research Paper.

\bibitem[{Amendola \textit{et~al.}(2019)Amendola, Candila and
  Gallo}]{Amendola:Candila:Gallo:2019}
Amendola, A., Candila, V. and Gallo, G.~M. (2019) On the asymmetric impact of
  macro--variables on volatility, \textit{Economic Modelling}, \textbf{76},
  135--152.

\bibitem[{Andersen and Bollerslev(1998)}]{Andersen:Bollerslev:1998}
Andersen, T.~G. and Bollerslev, T. (1998) Answering the skeptics: Yes, standard
  volatility models do provide accurate forecasts, \textit{International
  Economic Review}, \textbf{39}, 885--905.

\bibitem[{Andersen \textit{et~al.}(2006)Andersen, Bollerslev, Christoffersen
  and Diebold}]{Andersen:Bollerslev:Christoffersen:Diebold:2006}
Andersen, T.~G., Bollerslev, T., Christoffersen, P.~F. and Diebold, F.~X.
  (2006) Volatility and correlation forecasting, in \textit{Handbook of
  Economic Forecasting} (Eds.) G.~Elliott, C.~W.~J. Granger and A.~Timmermann,
  North Holland.

\bibitem[{Barigozzi \textit{et~al.}(2014)Barigozzi, Brownlees, Gallo and
  Veredas}]{Barigozzi+Brownless+Gallo+Veredas:2014}
Barigozzi, M., Brownlees, C., Gallo, G.~M. and Veredas, D. (2014) Disentangling
  systematic and idiosyncratic dynamics in panels of volatility measures,
  \textit{Journal of Econometrics}, \textbf{182}, 364--384.

\bibitem[{Barndorff-Nielsen \textit{et~al.}(2008)Barndorff-Nielsen, Hansen,
  Lunde and Shephard}]{BarndorffNielsen:Hansen:Lunde:Shephard:2008}
Barndorff-Nielsen, O.~E., Hansen, P.~R., Lunde, A. and Shephard, N. (2008)
  Designing realised kernels to measure the ex-post variation of equity prices
  in the presence of noise, \textit{Econometrica}, \textbf{76}, 1481--1536.

\bibitem[{Barndorff-Nielsen \textit{et~al.}(2009)Barndorff-Nielsen, Hansen,
  Lunde and Shephard}]{BarndorffNielsen:Hansen:Lunde:Shephard:2009}
Barndorff-Nielsen, O.~E., Hansen, P.~R., Lunde, A. and Shephard, N. (2009)
  Realised kernels in practice: trades and quotes, \textit{Econometrics
  Journal}, \textbf{12}, 1--32.

\bibitem[{Bollerslev(1986)}]{Bollerslev:1986}
Bollerslev, T. (1986) Generalized autoregressive conditional
  heteroskedasticity, \textit{Journal of Econometrics}, \textbf{31}, 307--327.

\bibitem[{Brownlees \textit{et~al.}(2011)Brownlees, Cipollini and
  Gallo}]{Brownlees:Cipollini:Gallo:2011}
Brownlees, C.~T., Cipollini, F. and Gallo, G.~M. (2011) Intra-daily volume
  modeling and prediction for algorithmic trading, \textit{Journal of Financial
  Econometrics}, \textbf{9}, 489--518.

\bibitem[{Brownlees \textit{et~al.}(2012)Brownlees, Cipollini and
  Gallo}]{Brownlees:Cipollini:Gallo:2012}
Brownlees, C.~T., Cipollini, F. and Gallo, G.~M. (2012) Multiplicative error
  models, in \textit{Volatility Models and Their Applications} (Eds.)
  L.~Bauwens, C.~Hafner and S.~Laurent, Wiley, pp. 223--247.

\bibitem[{Brownlees and Gallo(2010)}]{Brownlees:Gallo:2010}
Brownlees, C.~T. and Gallo, G.~M. (2010) Comparison of volatility measures: a
  risk management perspective, \textit{Journal of Financial Econometrics},
  \textbf{8}, 29--56.

\bibitem[{Cattivelli and Gallo(2020)}]{Cattivelli:Gallo:2020}
Cattivelli, L. and Gallo, G.~M. (2020) Adaptive lasso for vector multiplicative
  error models, \textit{Quantitative Finance}, \textbf{20}, 255--274.

\bibitem[{Cipollini \textit{et~al.}(2020)Cipollini, Gallo and
  Otranto}]{Cipollini:Gallo:Otranto:2020}
Cipollini, F., Gallo, G.~M. and Otranto, E. (2020) Realized volatility
  forecasting: Robustness to measurement errors, \textit{International Journal
  of Forecasting}, p. forthcoming.

\bibitem[{Conrad and Kleen(2020)}]{Conrad:Kleen:2020}
Conrad, C. and Kleen, O. (2020) Two are better than one: Volatility forecasting
  using multiplicative component {GARCH-MIDAS} models, \textit{Journal of
  Applied Econometrics}, \textbf{35}, 19--45.

\bibitem[{Conrad and Loch(2015)}]{Conrad:Loch:2015}
Conrad, C. and Loch, K. (2015) Anticipating long-term stock market volatility,
  \textit{Journal of Applied Econometrics}, \textbf{30}, 1090--1114.

\bibitem[{Corsi(2009)}]{Corsi:2009}
Corsi, F. (2009) A simple approximate long-memory model of realized volatility,
  \textit{Journal of Financial Econometrics}, \textbf{7}, 174--196.

\bibitem[{Dueker(1997)}]{Dueker:1997}
Dueker, M.~J. (1997) Markov switching in {GARCH} processes and mean-reverting
  stock-market volatility, \textit{Journal of Business \& Economic Statistics},
  \textbf{15}, 26--34.

\bibitem[{Engle(1982)}]{Engle:1982}
Engle, R.~F. (1982) Autoregressive conditional heteroscedasticity with
  estimates of the variance of {United Kingdom} inflation,
  \textit{Econometrica}, \textbf{50}, 987--1007.

\bibitem[{Engle(2002)}]{Engle:2002}
Engle, R.~F. (2002) New frontiers for {ARCH} models, \textit{Journal of Applied
  Econometrics}, \textbf{17}, 425--446.

\bibitem[{Engle and Gallo(2006)}]{Engle:Gallo:2006}
Engle, R.~F. and Gallo, G.~M. (2006) A multiple indicators model for volatility
  using intra-daily data, \textit{Journal of Econometrics}, \textbf{131},
  3--27.

\bibitem[{Engle \textit{et~al.}(2013)Engle, Ghysels and
  Sohn}]{Engle:Ghysels:Sohn:2013}
Engle, R.~F., Ghysels, E. and Sohn, B. (2013) Stock market volatility and
  macroeconomic fundamentals, \textit{Review of Economics and Statistics},
  \textbf{95}, 776--797.

\bibitem[{Engle and Rangel(2008)}]{Engle:Rangel:2008}
Engle, R.~F. and Rangel, J.~G. (2008) The spline-{GARCH} model for low
  frequency volatility and its global macroeconomic causes, \textit{Review of
  Financial Studies}, \textbf{21}, 1187--1222.

\bibitem[{Engle and Russell(1998)}]{Engle:Russell:1998}
Engle, R.~F. and Russell, J.~R. (1998) Autoregressive conditional duration: A
  new model for irregularly spaced transaction data., \textit{Econometrica},
  \textbf{66}, 1127--62.

\bibitem[{Gallo and Otranto(2015)}]{Gallo:Otranto:2015}
Gallo, G.~M. and Otranto, E. (2015) Forecasting realized volatility with
  changing average levels, \textit{International Journal of Forecasting},
  \textbf{31}, 620--634.

\bibitem[{Ghysels \textit{et~al.}(2006)Ghysels, Santa-Clara and
  Valkanov}]{Ghysels:SantaClara:Valkanov:2006}
Ghysels, E., Santa-Clara, P. and Valkanov, R. (2006) Predicting volatility:
  getting the most out of return data sampled at different frequencies,
  \textit{Journal of Econometrics}, \textbf{131}, 59--95.

\bibitem[{Ghysels \textit{et~al.}(2007)Ghysels, Sinko and
  Valkanov}]{Ghysels:Sinko:Valkanov:2007}
Ghysels, E., Sinko, A. and Valkanov, R. (2007) {MIDAS} regressions: Further
  results and new directions, \textit{Econometric Reviews}, \textbf{26},
  53--90.

\bibitem[{Glosten \textit{et~al.}(1993)Glosten, Jagannanthan and
  Runkle}]{Glosten:Jaganathan:Runkle:1993}
Glosten, L.~R., Jagannanthan, R. and Runkle, D.~E. (1993) On the relation
  between the expected value and the volatility of the nominal excess return on
  stocks, \textit{The Journal of Finance}, \textbf{48}, 1779--1801.

\bibitem[{Hamilton and Susmel(1994)}]{Hamilton:Susmel:1994}
Hamilton, J.~D. and Susmel, R. (1994) Autoregressive conditional
  heteroskedasticity and changes in regime, \textit{Journal of econometrics},
  \textbf{64}, 307--333.

\bibitem[{Han and Kristensen(2014)}]{Han:Kristensen:2014}
Han, H. and Kristensen, D. (2014) Asymptotic theory for the qmle in garch-x
  models with stationary and nonstationary covariates, \textit{Journal of
  Business \& Economic Statistics}, \textbf{32}, 416--429.

\bibitem[{Hansen \textit{et~al.}(2012)Hansen, Huang and
  Shek}]{Hansen:Huang:Shek:2012}
Hansen, P.~R., Huang, Z. and Shek, H.~H. (2012) Realized {GARCH}: a joint model
  for returns and realized measures of volatility, \textit{Journal of Applied
  Econometrics}, \textbf{27}, 877--906.

\bibitem[{Hansen \textit{et~al.}(2011)Hansen, Lunde and
  Nason}]{Hansen:Lunde:Nason:2011}
Hansen, P.~R., Lunde, A. and Nason, J.~M. (2011) The {M}odel {C}onfidence
  {S}et, \textit{Econometrica}, \textbf{79}, 453--497.

\bibitem[{Heber \textit{et~al.}(2009)Heber, Lunde, Shephard and
  Sheppard}]{Heber2009}
Heber, G., Lunde, A., Shephard, N. and Sheppard, K. (2009) Omi's realised
  library, version 0.1, Tech. rep., Oxford--Man Institute, University of
  Oxford.

\bibitem[{Ljung and Box(1978)}]{Ljung:Box:1978}
Ljung, G.~M. and Box, G. E.~P. (1978) On a measure of lack of fit in time
  series models, \textit{Biometrika}, \textbf{65}, 297--303.

\bibitem[{Mazur and Pipie{\'n}(2012)}]{Mazur:Pipien:2012}
Mazur, B. and Pipie{\'n}, M. (2012) On the empirical importance of periodicity
  in the volatility of financial returns-time varying {GARCH} as a second order
  {APC}(2) process, \textit{Central European Journal of Economic Modelling and
  Econometrics}, \textbf{2}, 95--116.

\bibitem[{Newey and McFadden(1994)}]{Newey:McFadden:1994}
Newey, W.~K. and McFadden, D. (1994) Large sample estimation and hypothesis
  testing, in \textit{Handbook of Econometrics} (Eds.) R.~F. Engle and
  D.~McFadden, Elsevier, vol.~4, chap.~36, pp. 2111--2245.

\bibitem[{Pan and Liu(2018)}]{Pan:Liu:2018}
Pan, Z. and Liu, L. (2018) Forecasting stock return volatility: A comparison
  between the roles of short-term and long-term leverage effects,
  \textit{Physica A: Statistical Mechanics and its Applications}, \textbf{492},
  168 -- 180.

\bibitem[{Patton(2011)}]{Patton:2011}
Patton, A. (2011) Volatility forecast comparison using imperfect volatility
  proxies, \textit{Journal of Econometrics}, \textbf{160}, 246--256.

\end{thebibliography}

\end{document}